%% file: main.tex
\documentclass[runningheads]{llncs}

\usepackage{eccv}

\usepackage{eccvabbrv}

\usepackage{graphicx}
\usepackage{booktabs}
\usepackage{comment}
\usepackage{amsmath,amssymb}
\input{preamble}

\usepackage[accsupp]{axessibility}  

\usepackage{hyperref}

\usepackage{orcidlink}

\begin{document}

\title{Depth from Dual Differential Defocus and Stereo (\dthreet) Consensus} 

\titlerunning{Abbreviated paper title}

\author{Junjie Luo\inst{1}\thanks{Equal contribution.} \and
Wei Xu\inst{1}\thanks{Equal contribution.} \and
Dylan Chu\inst{1} \and
Emma Alexander\inst{2} \and
Qi Guo\inst{1}}
 
\authorrunning{J.~Luo et al.}

\institute{Purdue University, West Lafayette, IN, USA \and
Northwestern University, Evanston, IL, USA}

\maketitle

\input{sec/0_Abstract}

\input{sec/1_Introduction}

\input{sec/2_Related_Work}

\input{sec/3_Methods}

\input{sec/4_Experiments}
\input{sec/5_Discussion}

\clearpage

\bibliographystyle{splncs}
\bibliography{egbib}

\clearpage

\setcounter{linenumber}{0}
\input{supp}

\end{document}

%% file: preamble.tex
\usepackage{lscape}
\usepackage{multirow}
\usepackage{makecell}
\usepackage{pifont}
\usepackage{lineno}

\newcommand{\va}{\boldsymbol{a}}

\newcommand{\vx}{\boldsymbol{x}}
\newcommand{\vZ}{\boldsymbol{Z}}
\newcommand{\vrho}{\boldsymbol{\rho}}

\newcommand{\x}{\boldsymbol{x}}

\newcommand{\X}{\boldsymbol{X}}

\newcommand{\dthree}{D$^3$}
\newcommand{\dthreet}{D$^3$S}

\newcommand{\citeS}[1]{\mbox{[{S}\citenum{#1}]}}

%% file: sec/0_Abstract.tex
\begin{abstract}
We introduce \dthreet~Consensus, a physics-based, closed-form algorithm that unifies depth-from-defocus (DfD) and stereo to achieve highly accurate depth estimation throughout an extended working range beyond the depth-of-field (DoF) of cameras. Given a pair of dual-defocus stereo images, the method estimates an overdetermined set of depth using a novel DfD theory, Dual Differential Defocus (\dthree), and (S)tereo in a coupled fashion. It then picks the most confident depth prediction from the set by enforcing consensus between these physically independent cues to reject unreliable estimates. \\

Analysis shows that \dthreet~achieves a comparable working range under the same error tolerance with 10$\times$ smaller baseline than previous triangulation-based depth estimation systems. This enables compact passive binocular rangefinders with substantially smaller form factors than conventional stereo and DfD designs. We demonstrate the first \dthreet~prototype with only 4~mm baseline and 12~mm EFL. It generates up to $900\times1800$-pixel depth maps with 1-cm mean absolute error over 0.3–1.64~m from a snapshot acquisition. This has surpassed the reported accuracy of certain commercially available stereo cameras with much larger form factors.
\end{abstract}

%% file: sec/1_Introduction.tex
\section{Introduction}

\begin{figure}[t]
  \centering
  \includegraphics[width=\linewidth]{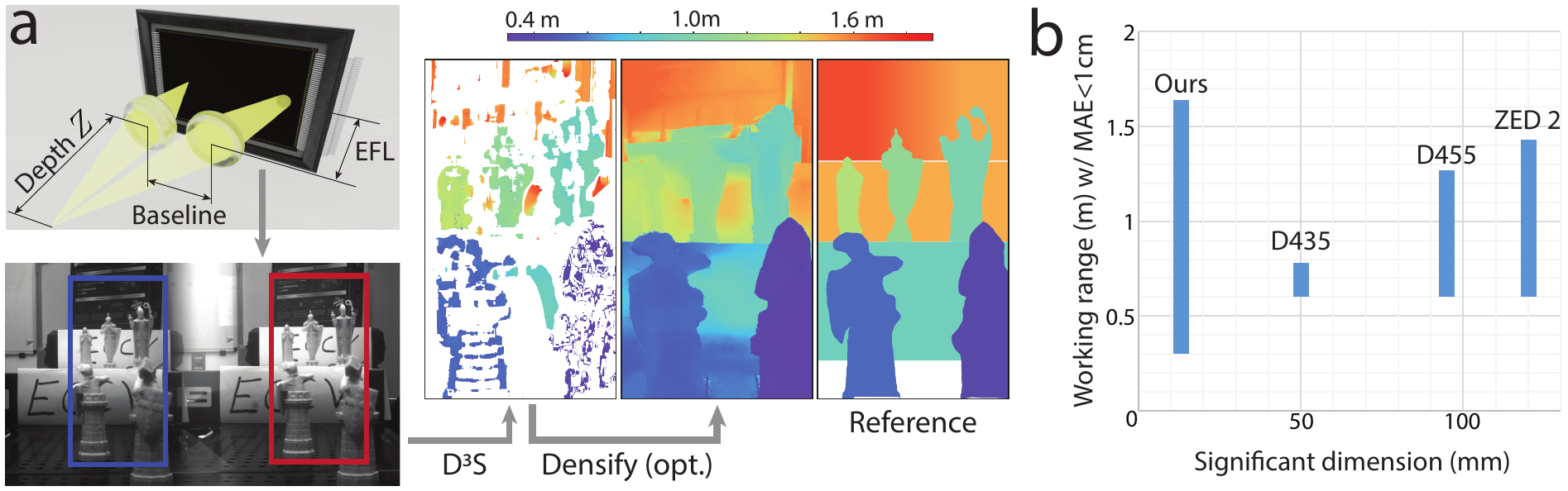}
  \caption{(a) Overview of \dthreet~Consensus. It utilizes a compact binocular camera system to capture two images of different defocus and perspectives (blue and red boxes). It then use the images to generate a highly accurate sparse depth map through a purely physics-based algorithm, which can be optionally densified using off-the-shelf depth completion models, e.g.~\cite{viola2025marigold}. (b) The accuracy of our \dthreet~prototype is sufficiently high to directly compare with commercially available stereo cameras. We estimate the working range with 1~cm MAE from community reported performances of several stereo camera products~\cite{rustler2025empirical} and compare them with ours. For fair comparison, these metrics are all measured from ranging front-parallel targets. The significant dimension is defined as the maximum between the baseline and the effective focal length (EFL), which reflects the compactness of the system. For all stereo methods, this is the baseline, while for us it is the EFL (12 mm, vs. a baseline of 4 mm). Ours demonstrates a clear advantage in both working range and compactness. }
  \label{fig:teaser}
\end{figure}

Triangulation is arguably the most widely used cue for passive depth estimation. However, its accuracy is theoretically bounded by the object's distance and the system's baseline and effective focal length (EFL)~(Fig.~\ref{fig:teaser}a)~\cite{schechner2000depth}. To achieve a working range, i.e., the range of depth under a pre-defined error tolerance, there exists a minimal length for the baseline or the EFL of the system, which in turn limits the system’s compactness. For example, for indoor-scale depth estimation, stereo cameras typically require at least 5-cm baseline~\cite{d435,d455,Stereolabs}, while depth-from-defocus (DfD) systems demand a similar EFL~\cite{guo2017focal,guo2019compact,luo2025focal,luo2025depth}. As a result, designing triangulation-based passive rangefinders that are both compact and capable of accurately sensing distant targets remains a longstanding challenge. 

We present \textit{\dthreet~Consensus}, a triangulation-based passive-ranging mechanism that pushes the trade-off limit between form factors and the working range. Given a pair of images with different perspectives and focus, \dthreet~Consensus derives depth from two intertwined but complementary cues. The first is a new image–depth relationship, \textit{Dual Differential Defocus (\dthree)}, which analytically links scene depth to the differential defocus observed when a camera undergoes two \textit{simultaneous} configuration changes. The second is the standard stereo equation. We show that by enforcing consensus between the \dthree- and stereo-based depth predictions, we obtain sparse yet substantially more accurate depth estimates than using either cue alone. This allows us to build a passive ranging system with a much smaller baseline and EFL than conventional ones, while still achieving on-par or even longer working range under uniform error tolerance~(Fig.~\ref{fig:teaser}b). 

The \dthree~relationship mathematically relates per-pixel scene depth $Z$ with the captured image $I$'s per-pixel spatial and optical derivatives:
\begin{align}
    Z_{\text{\dthree}}(\x) = \frac{\va^\top I_{\vx}(\x) + b~\lVert I_{\vx\vx}(\x) \rVert}{\dot{I}(\x) + c ~\lVert I_{\vx\vx}(\x) \rVert},
    \label{eq:d3_intro}
\end{align}
where $\x$ indicates the spatial coordinate of the image, $(I_{\vx}, I_{\vx\vx})$ are its per-pixel Jacobian and Hessian, $\lVert \cdot \rVert$ indicates the matrix norm, and $\dot{I}$ is the magnification-corrected, per-pixel intensity change with respect to optical parameters' variation. The coefficients $(\va, b, c)$ are determined by the specifications of the camera.

Consider a binocular camera system capturing a pair of stereo rectified images of a target, $(I_0, I_1)$, with slightly different focuses. \dthreet~Consensus evaluates Eq.~\ref{eq:d3_intro} by assuming the target to be at a set of candidate depths $\{Z^i\}_{i=1}^N$. This requires recalculating the image derivatives and coefficients in Eq.~\ref{eq:d3_intro} according to the stereo triangulation and the candidate depth value $Z^i$. If the estimated depth at a pixel $\x$, $Z_{\text{\dthree}}(\x)$, closely matches the candidate depth value $Z^i$, a consensus is established and the output depth value for position $\x$ is determined.  

We build a \dthreet~Consensus prototype, with a baseline of only 4~mm and an EFL of 12~mm. It generates an accurate, sparse depth map from a snapshot acquisition. In a thorough comparison with previous DfD and stereo systems, 
ours demonstrates on-par working range under uniform error tolerance, while only using a 10$\times$ smaller baseline. Its accuracy and working range is already comparable with commercially available stereo cameras while being much more compact (Fig.~\ref{fig:teaser}b).  We also demonstrate that the sparse depth maps can be optionally fed into off-the-shelf, pre-trained depth completion models~\cite{viola2025marigold} to produce densified depth maps (Fig.~\ref{fig:teaser}a).

The major contributions of this paper include the following:
\begin{enumerate}
    \item A closed-form DfD estimation theory, \dthree, that derives per-pixel depth maps from magnification-corrected, simultaneous differential defocus.
    \item A passive ranging mechanism, \dthreet~Consensus, that measures highly accurate depth in an extended range while utilizing a compact imaging platform.
    \item A prototype rangefinder using \dthreet~Consensus that demonstrates drastic improvements in both accuracy and compactness compared to previous works.
\end{enumerate}

%% file: sec/2_Related_Work.tex
\section{Related Work}




Estimating scene depth via triangulation is typically performed in two stages. The first stage extracts geometric cues—such as disparity in stereo or defocus in DfD—from intensity variations in the captured images~\cite{hirschmuller2005accurate, joshi2014micro,alexander2019theory,xu2025blurry}. The second stage densifies and refines the resulting depth map, particularly in textureless regions and near object boundaries, e.g. by leveraging high-level semantic priors~\cite{levin2007image,barron2016fast,tang2017depth,holynski2018fast}. Substantial progress has been made on this refinement stage using deep learning architectures~\cite{ma2018sparse,imran2019depth,tang2020learning,viola2025marigold}, and in many works, both stages are jointly performed within a single end-to-end model~\cite{tan20213d,wu2019phasecam3d,chang2019deep}. Complementary to these efforts, this paper concentrates on improving the geometric estimation stage itself. Accordingly, our review emphasizes prior work on the physical mapping from triangulation cues to depth.

Depth-from-defocus (DfD) estimates scene depths by analyzing the blurriness within small image patches~\cite{pentland1987new}. A substantial body of work has developed computational models capable of recovering depth maps from shallow depth-of-field (DoF) photos taken by commercial cameras~\cite{gur2019single,zhang2021joint,si2023fully,anwar2021deblur,nazir2023depth}. In parallel, researchers have explored co-designing specialized point spread functions (PSFs) to facilitate depth recovering, using engineered optical elements like coded apertures~\cite{levin2007image,zhou2009coded}, diffractive optical elements~\cite{wu2019phasecam3d,chang2019deep,ikoma2021depth}, and metasurfaces~\cite{hazineh2022d}. Another line of research exploits differential relationships between defocus and scene depths~\cite{farid1998range,subbarao1994depth,guo2017focal,alexander2018focal,guo2019compact,alexander2019theory,luo2025depth,luo2025focal,Ferreira2026SpiderCam}. Such methods can achieve significantly higher computational efficiency compared to conventional DfD frameworks.


DfD systems are typically monocular, with an effective baseline determined by the entrance pupil diameter—substantially smaller than the physical baseline of stereo configurations. As a result, achieving a comparable working range under the same error tolerance generally requires a longer effective focal length (EFL), which in turn limits system compactness. See \cite{schechner2000depth} for a thorough theoretical treatment of these trade-offs. Moreover, many DfD algorithms rely on deconvolution or differential operations, both of which amplify noise~\cite{takemura2019depth,luo2025depth}. This sensitivity to noise has further limited the feasible accuracy of DfD prototypes compared to stereo systems.

The working range of stereo systems is jointly constrained by the physical baseline and the DoF~\cite{zhang2014depth}, since accurate disparity estimation becomes difficult when image textures fall outside the DoF and appear blurred. To mitigate this limitation, prior work has explored extending the DoF using coded apertures~\cite{tan2021codedstereo,lopez2024low,takeda2012coded}, assigning different focal settings to the two cameras~\cite{li2010dual,takeda2013fusing}, or combining both strategies~\cite{takeda2013fusing,liu2025learned,ou2025learning}. Some recent learning-based approaches implicitly integrate defocus cues with stereo disparity estimation within a unified model~\cite{liu2025learned,ou2025learning}. However, to the best of our knowledge, physics-based methods that explicitly and analytically unify DfD and stereo within a single system have not yet been demonstrated.

%% file: sec/3_Methods.tex
\section{Dual Differential Defocus (\dthree) Theory}
\begin{figure}[t]
    \centering
    \includegraphics[width=0.7\linewidth]{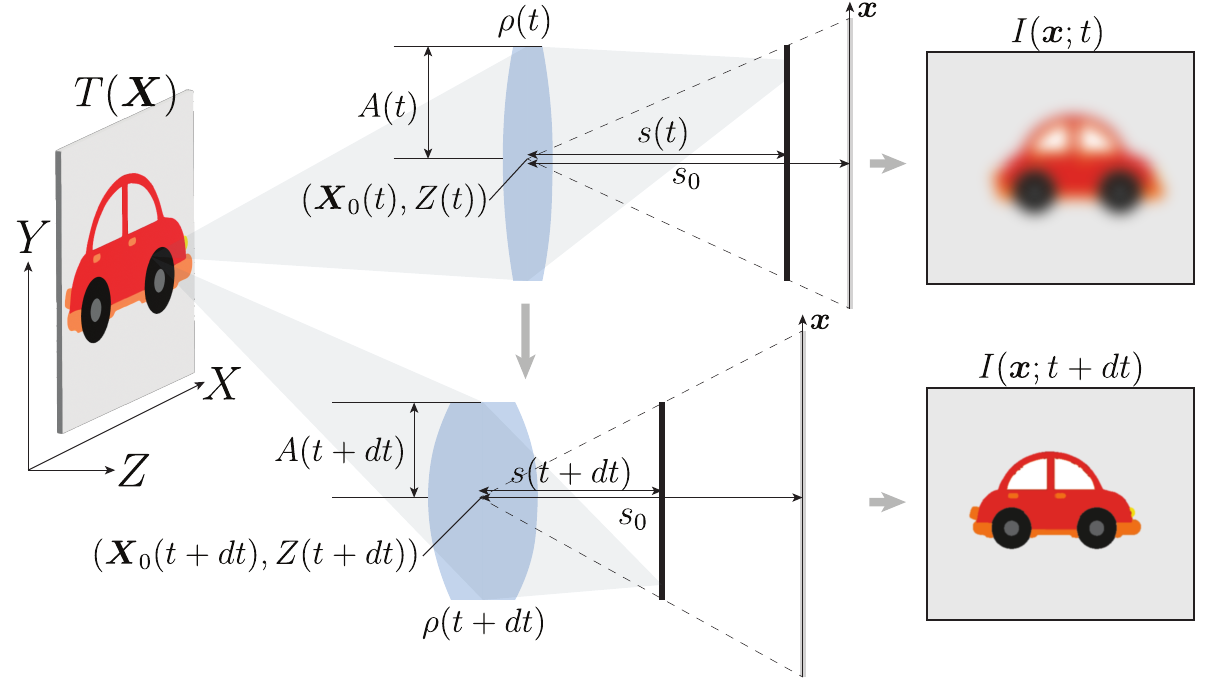}
    \caption{Image formation model. A front-parallel target with texture \(T(\x)\) is imaged by a thin-lens camera. This paper derives an analytical relationship between the image derivatives and target's depth from the camera $Z$ when some or all of the optical parameters, including optical power \(\rho\), aperture radius \(A\), sensor distance \(s\), and camera-center position \((\x_0, z)\) are differentially varied. }
    \label{fig:theory}
\end{figure}

\paragraph{Image Formation Model.} Consider a front-parallel target with texture $T(\X)$, $\X=(X,Y)$, imaged by a thin-lens camera with variable optical power $\rho$, aperture radius $A$, sensor distance $s$, and camera center position $(\X_0, Z)$, as shown in Fig.~\ref{fig:theory}. The image formed on the photosensor at time $t$, scaled to a standardized sensor distance $s_0$ and pixel pitch $p$ to isolate magnification, is:
\begin{equation}
    \begin{aligned}
    I(\x; t) =  k\left(\x; \sigma(t)\right)* 
    T\left(\X_0(t)-\frac{z(t)p}{s_0}\x\right),
    \label{eq:form}
\end{aligned}
\end{equation}
In this paper, we only consider single-lens cameras, therefore the term EFL and sensor distance is used interchangeably. 
As noted in Fig.~\ref{fig:theory}, we use $(\X,Z)$ to represent the world coordinate, and $\x$ to denote the photosensor coordinate. According to the thin-lens law, the point spread function (PSF) of the camera, $k(\x)$, takes the form:
\begin{align}
    k(\x; \sigma(t)) = \frac{1}{\sigma(t)^2}\kappa\left(\frac{\x}{\sigma(t)}\right),
\end{align}
where the term $\sigma$ is the \textit{defocus level} determined by the target's depth and optical parameters of the system:
\begin{align}
    \sigma(t)=\frac{s_0 A(t)}{s(t)}\left[\left(\frac{1}{Z(t)}-\rho(t)\right)s(t)+1\right],
\end{align}
and $\kappa(\x)$ is the pupil function that describes the transmittance profile of the aperture. For simplicity of notation, we omit the time variable $t$ when unambiguous. 

\paragraph{General Depth Relationship.} When the pupil function $\kappa(\x)$ can be approximated as a 2D Gaussian distribution: 
\begin{align}
    \kappa(\x) = \exp\left(-\frac{1}{2}\x^\top \Sigma^{-1} \x\right),
\end{align}
where $\Sigma \in \mathbb{R}^{2\times 2}$, the partial derivatives of the PSF are analytically related~\cite{guo2017focal}:
\begin{align}
    \dot{k}(\x) =\dot{\sigma}\sigma|\Sigma|^2A^2\nabla^2_{\Sigma}k(\x).
    \label{eq:dfdd}
\end{align} 
The overdot notation, $\dot{\Box}$, represents the derivative in $t$. The operator $\nabla^2_\Sigma \triangleq \nabla^T\Sigma \nabla$, where $\nabla = [\partial_x, \partial_y]^\top$. In the general case where the camera's optical power $\rho$, aperture radius $A$, sensor distance $s$, and center position $(\X_0,Z)$ are varied \textit{simultaneously} over a differential change of time $t$, the derivative of the defocus level, $\dot{\sigma}$, takes the form:
\begin{align}
    \dot{\sigma} = -s_0A\dot{\rho} -\frac{\dot{s}}{s^2} + s_0\left(\frac{1}{Z}-\rho\right)\dot{A} - \frac{s_0 A \dot{Z}}{Z^2}.
    \label{eq:dsigma}
\end{align}
According to Eq.~\ref{eq:form}, the image derivative with respect to the multiple, simultaneous variation of the optical parameters is:
\begin{equation}
    \begin{aligned}
    \dot{I}(\x)  =&\dot{k}\left(\x\right)* T\left(\X_0(t)-\frac{z(t)p}{s_0}\x\right) \\
    & \quad +k\left(\x\right)* \dot{T}\left(\X_0(t)-\frac{z(t)p}{s_0}\x\right) \\
     =\dot{\sigma}\sigma&|\Sigma|^2A^2\nabla^2_{\Sigma}I\left(\x\right) +  I_{\x}(\x)^\top \dot{\X}_0 - \frac{pI_{\x} (\x)^\top \x}{s_0}.
    \label{eq:derivative}
\end{aligned}
\end{equation}
Substituting Eq.~\ref{eq:dsigma} into the previous expression yields a cubic polynomial in the target depth \(Z\):
\begin{align}
a_3Z^3 + a_2Z^2 + a_1Z + a_0 = 0,
\label{eq:general}
\end{align}
where the coefficients \(a_{0:3}\) depend on the image derivatives \((\dot{I}, \nabla^2_\Sigma I, I_{\x})\) and the optical parameters and their temporal derivatives \((\rho, A, s, \X_0, \dot{\rho}, \dot{A}, \dot{\X}_0, \dot{Z})\), which are known and fixed for a given camera configuration. The explicit expressions of these coefficients are provided in the supplement. For clarity, the spatial variable \(\x\) is omitted, as the relationship applies independently to each pixel.

Eq.~\ref{eq:derivative} establishes a unified relationship between the target's depth $z$ and image derivatives with respect to all optical changes. It encompasses the depth estimation theories utilized in a series of prior depth-from-differential-defocus (DfDD) works~\cite{alexander2016focal, guo2017focal, alexander2018focal, guo2019compact, luo2025focal}, which leverage the variation of a single optical parameter, and complements Alexander~\cite{alexander2019theory} with an alternative formulation of the general DfDD equation that cancels the spatial scaling factor $s/s_0$. 

\paragraph{Dual Differential Defocus (\dthree).}  Eq.~\ref{eq:general} yields a third-order polynomial in depth $Z$, which may admit multiple solutions. In this paper, we focus on situations where the lateral position of the camera center $\X_0$ and an additional optical parameter, the lens' optical power $\rho$ or the sensor distance $s$, are varied. This leads to the following, simplified equation with a single solution: 
\begin{equation}
    Z_{\text{\dthree}}=\frac{ s_0 \, I_{\x}^\top \dot{\X}_0-\alpha\nabla^{2}_{\Sigma} I}{  \dot{I}-\alpha\left(\rho-1/s\right)\nabla^{2}_{\Sigma} I},
    \label{eq:d3}
\end{equation}
where $\alpha = |\Sigma|^{2}s_0^{2}A^{2}\dot{\rho}$ when the optical power $\rho$ is varied, and $\alpha = |\Sigma|^{2}A^{2}\dot{s}$ when the sensor distance $s$ is varied.

\paragraph{Snapshot Acquisition.}
Eq.~\ref{eq:d3} can be evaluated from a pair of images $(I_0, I_1)$ simultaneously captured by two cameras with slightly offset camera centers and optical powers:
\[
I_0 = I\!\left(\X_0 + \frac{\Delta \X}{2}, \rho + \frac{\Delta \rho}{2}\right), 
\quad
I_1 = I\!\left(\X_0 - \frac{\Delta \X}{2}, \rho - \frac{\Delta \rho}{2}\right).
\]
The required derivatives can be approximated using finite differences:
\begin{align}
    \dot{I} &\approx I_0 - I_1, \nonumber \\
    I_{\x} &\approx \nabla (I_0 + I_1)/2, \nonumber \\
    \nabla^2_\Sigma I &\approx \nabla^2_\Sigma (I_0 + I_1)/2.
    \label{eq:finite-diff}
\end{align}
Such acquisition can be realized using a beamsplitter-based setup, e.g.~\cite{Ferreira2026SpiderCam}, if \dthree{} is used as the sole depth cue. In this work, however, we combine the \dthree{} cue with stereo by leveraging snapshot measurements from a compact binocular system together with the novel algorithm described in Sec.~\ref{sec:d3t}.

\section{\dthreet~Consensus}
\label{sec:d3t}

\begin{figure}[t]
    \centering
    \includegraphics[width=0.7\linewidth]{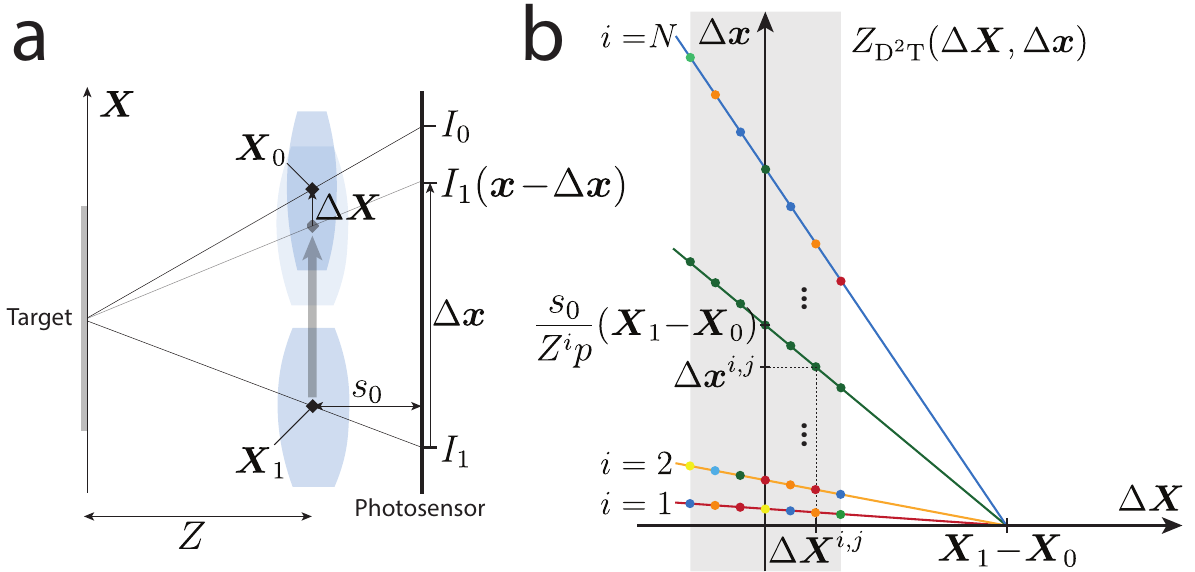}
    \caption{Overview of \dthreet~Consensus. (a) Consider two cameras centered at $\X_0$ and $\X_1$ with different optical powers imaging a front-parallel target at depth $Z$. We shift the captured images $I_1$ towards $I_0$ by $\Delta \x$ so that the effective baseline distance between $I_0$ and the shifted $I_1$, $\Delta \X$, is sufficiently small for accurately computing Eq.~\ref{eq:consensus} via finite difference approximation. (b) To solve the target depth, \dthreet~Consensus generates a dense set of depth predictions using Eq.~\ref{eq:consensus} under different combinations of $\Delta \X$ and $\Delta \x$. See the colored points where colors denote estimated depth values. The true depth can be determined based on the novel consensus criteria, where the depth predictions of points along a line must be consistent with each other and with the depth value indicated by the line. }
    \label{fig:consensus}
\end{figure}

Consider having two cameras (0 and 1) with lens centers located at $(\X_0, Z)$ and $(\X_1, Z)$, as shown in Fig.~\ref{fig:consensus}a. Both cameras share the same aperture radius $A$ and sensor distance $s_0$, while their optical powers are $\rho+\Delta\rho$ and $\rho-\Delta\rho$, respectively. The minimal baseline separation between the cameras, $\lVert \X_0 - \X_1\rVert$, must be larger than twice the aperture radius, $2A$, to avoid collision of the two cameras. Consequently, the disparity between the captured images $I_0$ and $I_1$ becomes significantly larger than the characteristic texture scale, causing the finite difference $I_0 - I_1$ to deviate from being a close approximation to $\dot{I}$. Therefore, the closed-form solution (Eq.~\ref{eq:d3}) cannot be directly applied. 

We propose the \textit{\dthreet~Consensus}, which generates a per-pixel depth map $Z$ from a pair of differently defocused images captured from a binocular system, $I_0$ and $I_1$. We provide a high-level summary followed by a more rigorous description of this algorithm below. 

Given a set of densely sampled candidate depth values $\{Z^i\}_{i=1}^N$, we translate $I_1$ towards $I_0$ by the disparity corresponding to each $Z^i$. Instead of directly evaluating the matching quality between $I_0$ and the translated $I_1$, we use \dthree~(Eq.~\ref{eq:d3})~to generate overdetermined per-pixel depth estimates. These estimates are visualized as colored points in Fig.~\ref{fig:consensus}b, where the color encodes the predicted depth value (the same convention applies below).

We then enforce a stereo-based consensus constraint. For each candidate depth $Z^i$, the corresponding \dthree~depth estimates fall on a line, with the color indicating the candidate depth value. When the candidate depth $Z^i$ is incorrect, the corresponding \dthree~depth estimates do not agree with the candidate depth value: As visualized in Fig.~\ref{fig:consensus}b, the colors of the points along a line are inconsistent and do not match that of the line. The correct candidate depth yields consistent colors across all points and the line. This novel \dthreet~Algorithm enforces consistency between two independent depth cues, the \dthree~(Eq.~\ref{eq:d3})~and Stereo, to filter out depth estimations contaminated by noise, leading to significantly higher depth estimation accuracy than using only one depth cue. 

\paragraph{Overdetermined Depth Estimation using \dthree.} Consider imaging a front-parallel target at depth $Z$ as visualized in Fig.~\ref{fig:consensus}a. By pixel shifting $I_1$ towards $I_0$ with displacement $\Delta \x$, the resulting image pair $(I_0, I_1(\x+\Delta \x))$ can be made equivalent to one captured from a hypothetical binocular system with the effective baseline distance $\Delta \X$, where this virtual baseline can allow the lenses to overlap as required by the differential shift assumption. For accurate reprojection through the virtually shifted lens, the image translation $\Delta \x$ depends on both the effective baseline $\Delta \X$ and the unknown depth $Z$ according to \textit{stereo}: 
\begin{align}
    \Delta \x = \frac{s_0}{Zp} \left( \X_0 - \X_1  - \Delta \X\right).
    \label{eq:line}
\end{align}

\noindent Because the target's depth $Z$ is unknown, the \dthreet~Consensus searches a range of candidate values $\{Z^i\}_{i=1}^N$. For each $Z^i$, it loops over a set of virtual baselines $\{\Delta \X^{i,j}\}_{j=1}^M$, each virtual baseline $\X^{i,j}$ corresponds to a shift $\Delta \x^{i,j}$ according to Eq.~\ref{eq:line}. With this shift, we can then solve for depth  
using Eq.~\ref{eq:d3} on the shifted image pair $(I_0, I_1(\x +\Delta\x^{i,j}))$ with the finite differences in Eq.~\ref{eq:finite-diff}. 
We denote the resulting depth values as $Z_{\text{\dthree}}(\Delta \X^{i,j}, \Delta \x^{i,j})$. This method provides $M$ depth predictions for each pixel at each candidate depth $Z^i$ using the \dthree~cue, visualized for $M=7$ as the colored points along each line in Fig.~\ref{fig:consensus}b.

\paragraph{Enforcing Consensus Between \dthree~and Stereo.} Now, we have two sources of depth estimation, the assumed $Z^i$ and the computed $Z_{\text{\dthree}}(\Delta \X^{i,j}, \Delta \x^{i,j})$. If the candidate depth $Z^i$ is the true target depth at a pixel and the equivalent baseline distance $\Delta \X^{i,j}$ is sufficiently small so that the finite difference approximation (Eq.~\ref{eq:finite-diff}) is accurate, it must satisfy the following criterion for \textit{consensus}:
\begin{equation}
    \begin{aligned}
    \left|Z^i - Z_{\text{\dthree}}(\Delta \X^{i,j}, \Delta \x^{i,j})\right| &< \Delta Z, \\
    \left|1/Z^i - 1/Z_{\text{\dthree}}(\Delta \X^{i,j}, \Delta \x^{i,j})\right| &< \Delta \rho, \; \forall j
    \label{eq:consensus}
\end{aligned}
\end{equation}
where $\Delta Z$ and $\Delta \rho$ are predetermined tolerance constants. 

Fig.~\ref{fig:consensus}b explains such consensus intuitively. The proposed algorithm creates a nonuniformly-sampled space of depth estimations $Z_{\text{\dthree}}(\Delta\X, \Delta\x)$ given different combinations of effective baseline and image translation $(\Delta\X, \Delta\x)$. The candidate depth $Z^i$ and the initial baseline $\X_0 - \X_1$ determine a line within this solution space as defined by Eq.~\ref{eq:line}. If the candidate depth $Z^i$ matches the ground truth and $\lVert\Delta\X^{i,j}\rVert$ is reasonably small, all depth solutions along the corresponding line, $Z_{\text{\dthree}}(\Delta\X^{i,j}, \Delta\x^{i,j})$, should be equal to each other. Furthermore, these predictions must equal the $Z^i$ indicated by the slope of the line. Consensus is shown by the green dots along the green line in Fig.~\ref{fig:consensus}b. 

We enforce consensus on both depth $Z$ and inverse depth $1/Z$ in Eq.~\ref{eq:consensus}. Empirically, we observed that enforcing consensus solely on $Z$ can lead to spurious agreement at small candidate depths $Z^i$, regardless of the true depth value. Incorporating consensus in the inverse-depth domain effectively suppresses these false positives. A more detailed analysis of this phenomenon is provided in the supplementary material.

\paragraph{Consensus as Confidence.} Compared to previous differential defocus algorithms~\cite{guo2017focal, guo2019compact, luo2025focal, luo2025depth} that rely on an additional confidence metric to determine the reliability of depth predictions, the consensus mechanism directly inspires a quantitative metric to filter out inaccurate depth predictions. For each candidate depth value $Z^i$ at each pixel $\x$,  we define its confidence value as:
\begin{align}
    C^i = \left[\left(1+\lVert \Delta \vZ^i \rVert_{\infty} \right)\left(1+\lVert \Delta \vrho^i \rVert_{\infty} \right)\right]^{-1},
    \label{eq:Ci}
\end{align}
where $\Delta\vZ^i$ and $\Delta\vrho^i$ are the non-consistency vectors in the depth and inverse depth domain for the candidate depth $Z^i$:
\begin{equation*}
    \begin{aligned}
    \Delta\vZ^i = \begin{bmatrix}
        Z^i - Z_{\text{\dthree}}(\Delta \X^{i,1}, \Delta \x^{i,1}) \\
        \vdots \\
        Z^i - Z_{\text{\dthree}}(\Delta \X^{i,M}, \Delta \x^{i,M})
    \end{bmatrix}, 
    \Delta\vrho^i = \begin{bmatrix}
        1/Z^i - 1/Z_{\text{\dthree}}(\Delta \X^{i,1}, \Delta \x^{i,1}) \\
        \vdots \\
        1/Z^i - 1/Z_{\text{\dthree}}(\Delta \X^{i,M}, \Delta \x^{i,M})
    \end{bmatrix}, 
\end{aligned}
\label{eq:conf}
\end{equation*}
For each pixel, we set the depth prediction to be the candidate depth $Z^i$ with the highest confidence value $C^i$:
\begin{align}
    Z = Z^{\arg\max_i C^i}, 
    C = \max_i C^i
\end{align}
and then filter the depth prediction with a predefined confidence threshold $C_\text{thre}$:
\begin{align}
    Z_{\text{filtered}} = \begin{cases}
        Z, & C > C_\text{thre}\\
        \text{No output,} & \text{otherwise}.
    \end{cases}
    \label{eq:threshold}
\end{align}
The output of the \dthreet~Algorithm is a sparse depth map with all depth predictions that satisfies the consensus between the \dthree~and triangulation cues.

%% file: sec/4_Experiments.tex
\section{Experimental Results}

\subsection{Prototype Camera System}
\label{secsec:prototype}

\begin{figure}[t]
    \centering
    \includegraphics[width=0.9\linewidth]{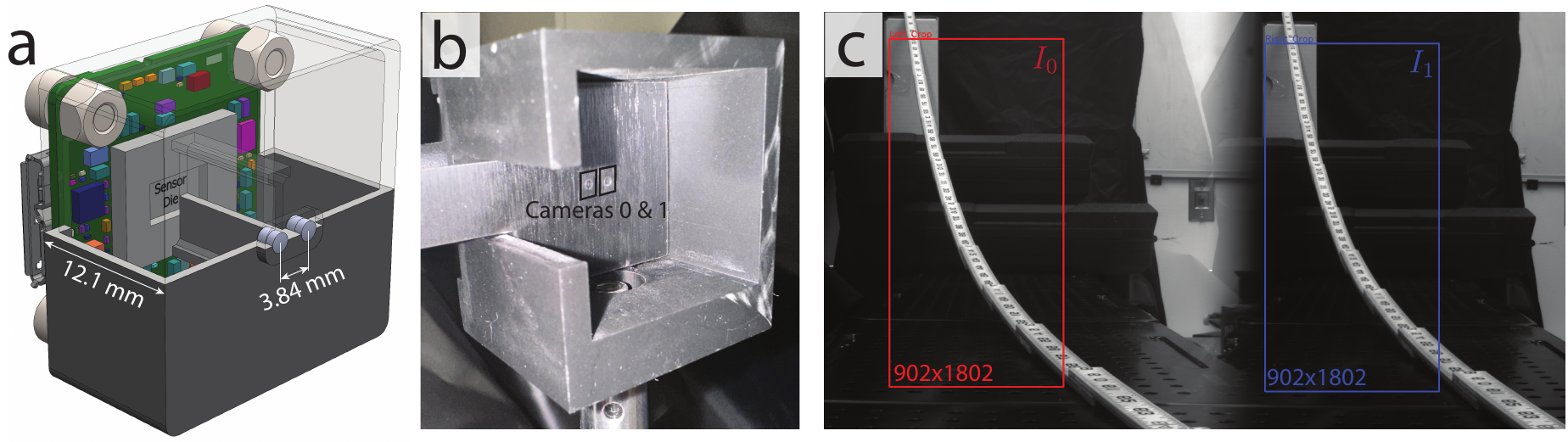}
    \caption{\dthreet{} Consensus Prototype. 
(a) CAD rendering of the system assembly. 
(b) Photograph of the built prototype. 
(c) Sample image pair captured in a laboratory environment. 
The raw measurement exhibits slight overlap between the two views; non-overlapping regions are cropped to obtain $I_0$ and $I_1$. 
A measuring tape is included in the scene to indicate the focal plane of each image. 
The images $I_0$ and $I_1$ contain both stereo disparity and a slight defocus difference between the two views. }
    \label{fig:prototype}
\end{figure}

We built a prototype imaging system to validate \dthreet{} Consensus, as shown in Fig.~\ref{fig:prototype}a-b. The system consists of two side-by-side cameras with off-the-shelf achromatic doublets (2\,mm aperture diameter, 12\,mm focal length) that form two side-by-side images $(I_0, I_1)$ on a shared image sensor. Custom 3D-printed holders and separators are designed to ensure mechanical alignment and minimize optical crosstalk between the two cameras. 

Due to the limited availability of off-the-shelf optics, we use identical lens models for both cameras while introducing a slight difference in sensor distance $s_0$ to generate the required defocus variation of Eq.~\ref{eq:d3}. We observe that the magnification change between $(I_0, I_1)$ are negligible, while the defocus difference is clear. Therefore, we approximately treat the two cameras as with uniform sensor distance $s_0$ and slightly different optical powers $\rho\pm\Delta\rho$.

The designed baseline of the system is 3.84\,mm. A representative raw measurement is shown in Fig.~\ref{fig:prototype}c, captured at a resolution of 3840$\times$2160 pixels. From this measurement, we crop two corresponding regions of with maximal dimensions 902$\times$1802 pixels to obtain the images $I_0$ and $I_1$, respectively. 

A complete list of parts, CAD models for the 3D-printed components, and the assembling procedures are provided in the supplementary material to facilitate reproducibility.



\subsection{Calibration and Implementation}

We used the \dthreet{} prototype to capture image pairs $\{(I_{0,j}, I_{1,j})\}_{j=1}^J$ of a fronto-parallel textured target placed at known distances $\{Z_j\}_{j=1}^J$. These measurements serve as calibration data for determining the parameters of \dthree{}. To avoid information leakage, this target is not used in any subsequent experiments or analyses.

For practical parameter tuning, we approximate and simplify Eq.~\ref{eq:d3} to the following form:
\begin{align}
Z_{\text{\dthree}}(\x) = \frac{a(\Delta\X) \sum_{\x'\in \mathcal{B}(\x)} I_x(\x')\dot{I}(\x')}{\sum_{\x'\in \mathcal{B}(\x)} \dot{I}^2(\x')} + b(\Delta\X),
\label{eq:d3_actual}
\end{align}
where $a(\Delta\X)$ and $b(\Delta\X)$ are computational parameters that depend on the equivalent baseline $\Delta\X$ defined in Sec.~\ref{sec:d3t}. The set $\mathcal{B}(\x)$ denotes the pixels within a window centered at pixel $\x$. The summations correspond to a least-squares solution evaluated over $\mathcal{B}(\x)$, which improves robustness to noise.

Using the collected calibration data, we estimate the parameters $a(\Delta\X)$ and $b(\Delta\X)$ and fix it throughout the real-world experiment. Additional details of the calibration procedure and the hyperparameter selections are provided in the supplementary material.

\subsection{Analysis of the Prototype}

\paragraph{Depth Accuracy.}
We evaluate the depth accuracy of the \dthreet{} prototype by estimating depth maps of four front-parallel planes textured with patterns randomly selected from the \textit{CUReT} dataset~\cite{dana1999curet}. The selected textures are shown in the supplementary material. Fig.~\ref{fig:accuracy}a plots the mean absolute error (MAE) of the predicted depth as a function of true depth under different confidence thresholds. Throughout the paper, we define the \emph{working range} as the continuous depth interval for which the MAE remains below 1~cm. As shown in Fig.~\ref{fig:accuracy}a, the proposed system achieves a working range from 0.3~m to 1.64~m when the confidence threshold is set to $C_\text{thre} = 0.8$.

\paragraph{Effect of Confidence.}
Fig.~\ref{fig:accuracy}b illustrates how the working range and the density of depth predictions vary with the confidence threshold $C_\text{thre}$. As the threshold increases from 0.5 to 0.8, the proportion of retained depth predictions decreases sharply, while the working range expands noticeably. This behavior indicates that the confidence threshold effectively filters unreliable estimates and extends the reliable operating range of the system.


\paragraph{Comparison of Different Methods.}
We process the same dataset using several representative baselines, including a previous DfD method~\cite{guo2017focal}, dual-defocus stereo approaches~\cite{li2010dual,takeda2013fusing}, and a simple baseline that estimates disparity by matching SIFT keypoints~\cite{lowe2004distinctive} between the two images. As in Fig.~\ref{fig:accuracy}c, all these methods exhibit substantially lower depth accuracy compared to \dthreet{}~Consensus.

\begin{figure}[t]
    \centering
    \includegraphics[width=\linewidth]{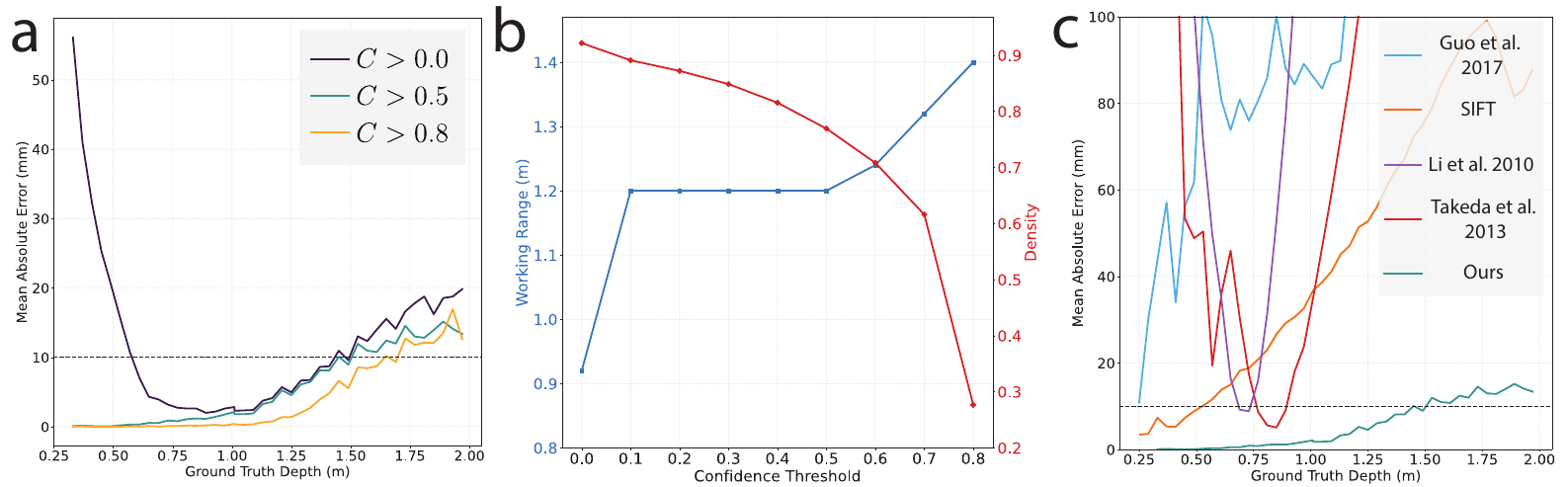}
    \caption{Analysis of the \dthreet{}~Consensus prototype in real experiments.
(a) Mean absolute error (MAE) as a function of ground-truth depth under different confidence thresholds. With a confidence threshold of $C_\text{thre}=0.8$, the prototype achieves a working range from 0.3~m to 1.64~m while maintaining MAE below 1~cm.
(b) Working range (blue) and prediction density (red) as functions of the confidence threshold. Increasing the threshold produces a two-stage effect. First, a small threshold (e.g., $0.1$) removes clear outliers, leading to a substantial increase in working range with minimal reduction in density. In the intermediate regime ($0.1$–$0.5$), the working range remains nearly constant. When the threshold becomes high ($>0.5$), the confidence criterion increasingly enforces agreement between the \dthree{} and stereo estimates, further extending the working range but at the cost of a noticeable reduction in prediction density. (c) Comparison of using different methods~\cite{guo2017focal,li2010dual,takeda2013fusing} to estimate depth maps from the same set of images captured by \dthreet{}~Consensus prototype.}
    \label{fig:accuracy}
\end{figure}

\subsection{Simulation Comparison with Previous Systems}

In this section, we compare \dthreet{} with prior DfD and dual-defocus stereo systems~\cite{takeda2012coded,takeda2013fusing,tang2017depth,tan2021codedstereo,liu2025learned,ou2025learning}. 
Because most of the compared works do not provide publicly available code or pre-trained models and report quantitative results primarily from simulation, we extract the reported metrics directly from the published papers and approximately convert them to mean absolute errors (MAEs) in the depth domain. 
For fairness, we additionally simulate a \dthreet{} system based on the prototype in Sec.~\ref{secsec:prototype} to perform the comparison.


For methods that report end-point errors (EPE) on disparity, $\epsilon_d$, over the full depth range of a benchmark dataset, we assume the error to be approximately constant across distances and convert it to a depth-domain MAE, $\epsilon_Z$, using a first-order Taylor approximation: 
\begin{align}
    \epsilon_Z = \epsilon_d Z^2 p/(s_0 \lVert\Delta \X\rVert),
    \label{eq:epe}
\end{align}
where $p$ denotes the pixel pitch. Because this conversion implies zero error at $Z=0$, it does not yield a meaningful lower bound on the working range; therefore, in our comparison we report only the upper bound derived from this conversion. 

For the simulated \dthreet{} system, we set the parameters to match those of the prototype: $\lVert \Delta \X \rVert = 3.84$~mm, $s_0 = 12.1$~mm, $\rho = 83.83~\text{m}^{-1}$, and $\Delta\rho = 0.05~\text{m}^{-1}$. The parameters in Eq.~\ref{eq:d3} can therefore be computed directly from these values. Unlike the real experiments, we use the original \dthree{} formulation (Eq.~\ref{eq:d3}) to compute $Z_{\text{\dthree}}$, since all required parameters are known from the simulation setup. We evaluate the simulated \dthreet{} system on a synthetic dataset containing realistic foreground and background textures, as well as scenes with complex depth discontinuities. Detailed descriptions of the synthetic dataset are provided in the supplement.

Figure~\ref{fig:system_comparison} plots the projected working ranges of different systems as a function of their baselines or effective focal lengths (EFLs). Our method occupies a distinct region of this design space, achieving a compelling combination of compactness and operating range. In particular, it breaks the traditional trade-off between working range and dimension, achieving a substantially smaller physical dimension than the other approaches while maintaining a comparable working range. The full specifications of these systems are listed in the supplement. 


We note that all compared systems produce dense depth maps, whereas our method outputs sparse depth estimates. Although dense outputs provide more spatially complete reconstructions, they do not necessarily yield higher depth accuracy at individual pixels. In many cases, the densification process mainly improves spatial coverage rather than the precision of the underlying depth measurements~\cite{tang2017depth, viola2025marigold}. Therefore, in this analysis we focus on comparing the intrinsic ranging accuracy of the systems.

It is important to emphasize that the compared approaches and ours target somewhat different operating points in the design space. Systems that produce dense depth maps are often optimized for applications requiring complete scene reconstruction, whereas our approach prioritizes compact hardware and accurate depth estimation over an extended working range. Consequently, the results presented here should be interpreted as a complementary reference point that highlights a compelling combination of accuracy and compact form factor, rather than as a general claim of superiority over existing dense-depth systems.

\begin{figure}[t]
    \centering
    \includegraphics[width=0.95\linewidth]{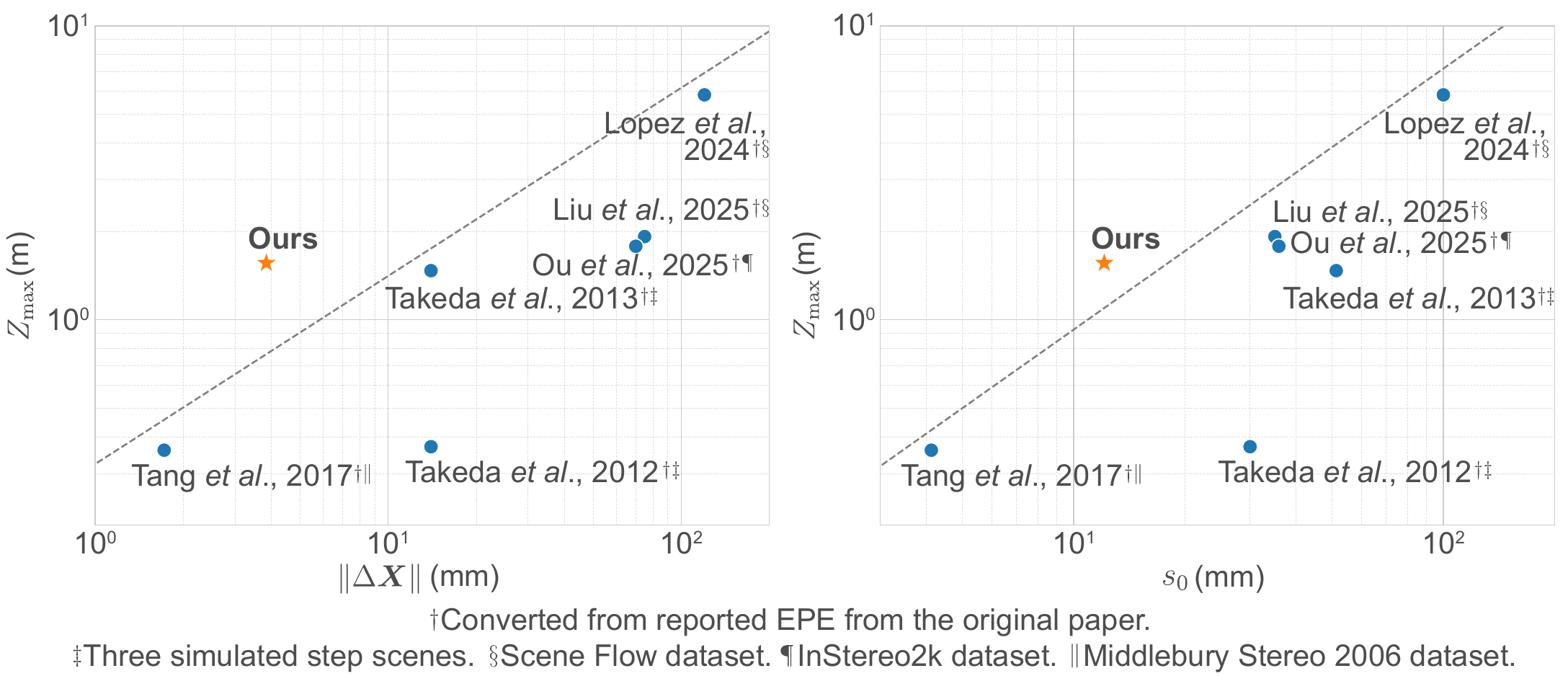}
    \caption{Projected working ranges of prior depth-from-defocus and dual-defocus stereo systems~\cite{takeda2012coded,takeda2013fusing,tang2017depth,tan2021codedstereo,liu2025learned,ou2025learning}, estimated from their reported simulation performance. The dashed line illustrates the traditional trade-off between system compactness and working range for triangulation-based rangefinders. Our method departs from this trend by achieving a substantially more compact form factor while maintaining a long projected working range. $Z_\text{max}$ denotes the upper bound of working range with MAE < 1 cm. $\lVert\Delta\X\rVert$ and $s_0$ represent the baseline and the EFL, respectively. }
    \label{fig:system_comparison}
\end{figure}

\subsection{Real World Depth Maps}

\begin{figure*}[t]
  \centering
    \includegraphics[width=\textwidth]{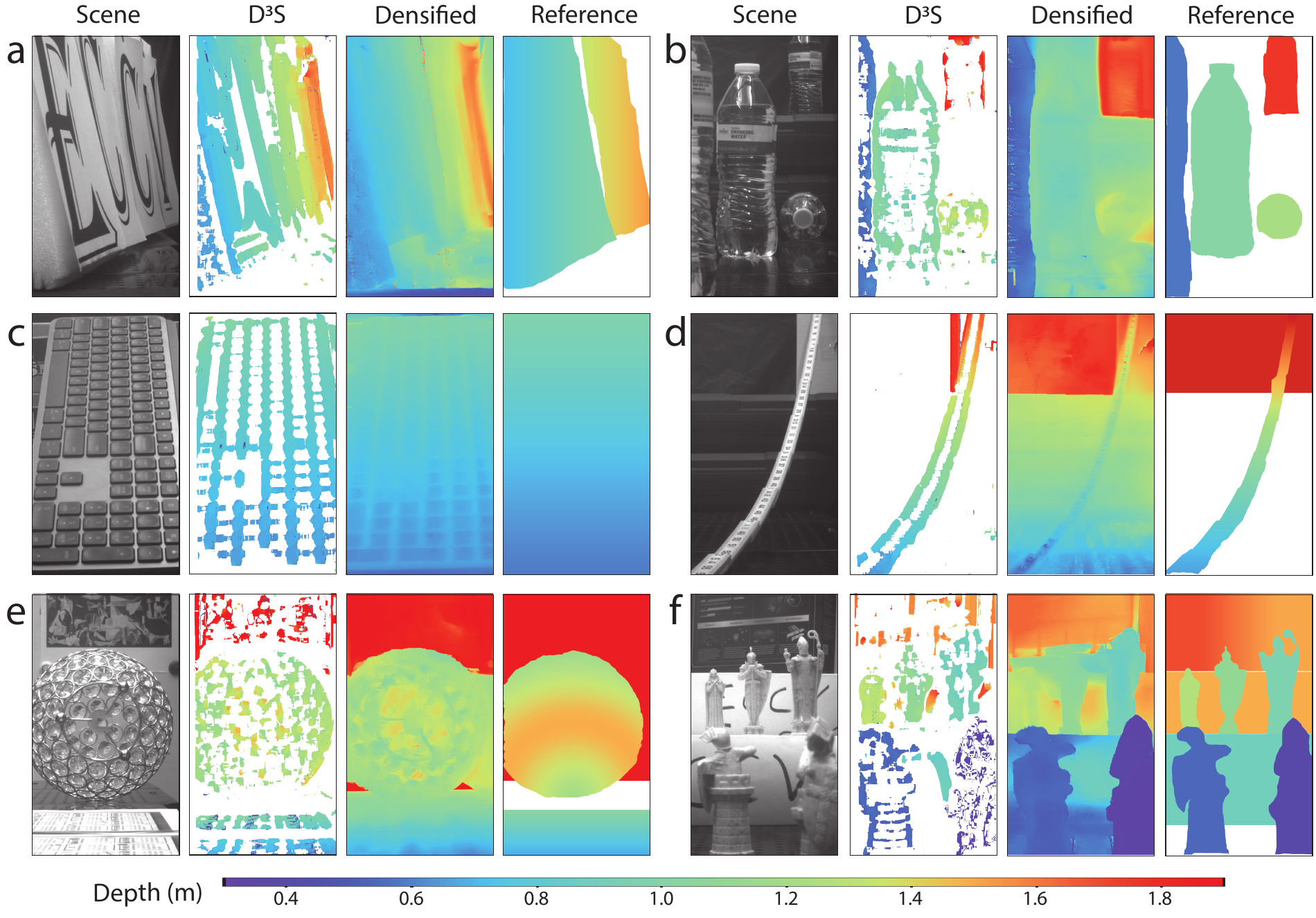}
  \caption{Sample real-world scenes captured by the \dthreet{} prototype. The sparse depth maps generated by \dthreet{} (second column) are filtered using Eq.~\ref{eq:threshold} with a constant threshold $C_{\text{thre}}=0.5$. These results demonstrate reliable depth estimation across diverse objects and surface textures. The densified depth maps (third column) are obtained by feeding the sparse depth maps together with the image $I_0$ into an off-the-shelf depth completion model~\cite{viola2025marigold}. Errors in the densified outputs are likely due to domain shift, as the model is directly applied to measurements from our prototype without fine-tuning. Reference depth maps are constructed from manual measurements.}
  \label{fig:real-world}
\end{figure*}

Fig.~\ref{fig:real-world} presents depth reconstructions from representative real-world scenes captured by the \dthreet{} prototype. The results demonstrate stable performance across diverse materials and reflectance properties, including moderately specular objects such as glasses and water bottles, over a working range of 0.4\,m to 1.8\,m. This suggests that \dthreet{} maintains reliable geometric estimation across a variety of surface characteristics.



As shown in Fig.~\ref{fig:real-world}c, a small number of pixels along the vertical edge of the keyboard exhibit abnormally near depth predictions. This behavior is likely caused by repetitive textures, which can produce multiple candidate depths satisfying the \dthreet{} consensus constraint. Such ambiguity is an inherent limitation of stereo-based triangulation methods.

Fig.~\ref{fig:real-world}e–f shows that \dthreet{} tends to suppress depth predictions near object boundaries. This is favorable as depth boundaries break the assumption of \dthreet{} and the predictions near these areas will be inaccurate. This behavior arises because pixels near depth discontinuities often contain mixed defocus and disparity cues from neighboring surfaces, which makes it difficult for the consensus mechanism to establish a reliable agreement between the two estimators.

For visualization purposes only, we further apply a pre-trained depth completion model, Marigold-DC~\cite{viola2025marigold}, to the sparse \dthreet{} outputs. Noticeable artifacts appear in textureless regions of the densified results. This behavior is likely due to distribution mismatch: Marigold-DC is trained primarily on LiDAR-style sparse inputs, where measurements are randomly distributed across both textured and textureless regions, whereas \dthreet{} produces sparse estimates concentrated in textured areas. Depth densification is beyond the scope of this work.

%% file: sec/5_Discussion.tex


%% file: supp.tex






\begin{center}
\textbf{\fontsize{14}{16.8}\selectfont Depth from Dual Differential Defocus and Stereo (\dthreet) Consensus}\\
\vspace{0.1in}
\fontsize{14}{16.8}\selectfont Supplementary Material
\vspace{0.1in}
\end{center}



\setcounter{section}{0}
\setcounter{figure}{0}
\setcounter{table}{0}
\setcounter{equation}{0}

\renewcommand{\thesection}{S\arabic{section}}
\renewcommand{\thefigure}{S\arabic{figure}}
\renewcommand{\thetable}{S\arabic{table}}
\renewcommand{\theequation}{S\arabic{equation}}

\section{Additional Information of \dthreet{} Consensus Algorithm}
\input{sec/Supp_full_experssion}

\subsection{Confidence}

\begin{figure}
    \centering
    \includegraphics[width=0.7\linewidth]{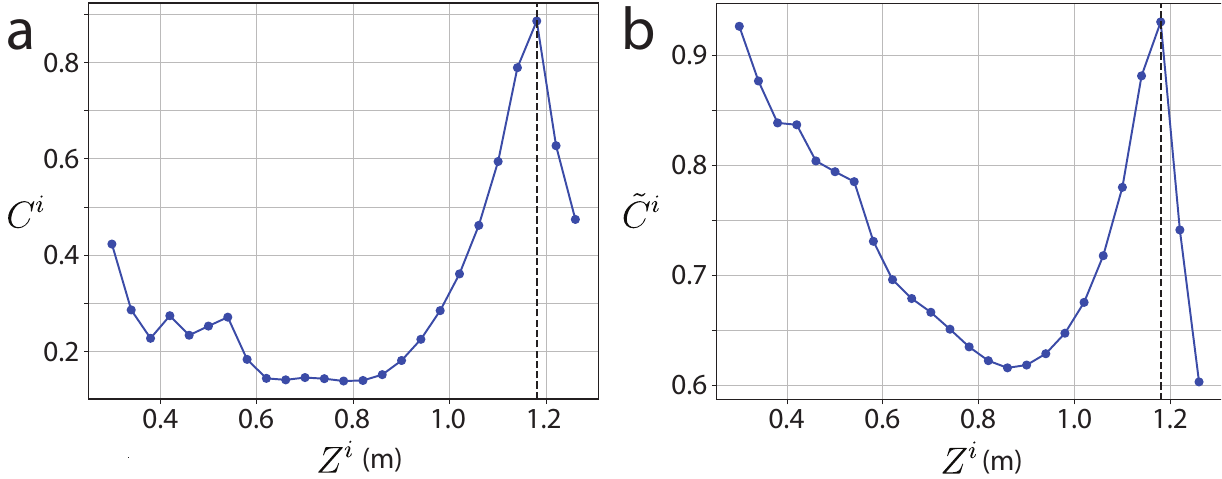}
    \caption{Confidence values at different candidate depths $Z^i$ calculated using (a) Eq.~\ref{eq:Ci} and (b) Eq.~\ref{eq:Ci_Z} using real-world data. The vertical dashed line indicate the true depth. }
    \label{fig:ZiCi}
\end{figure}

Figure~\ref{fig:ZiCi} shows the confidence values corresponding to different candidate depths $Z^i$ at a representative pixel from real-world data. The values are computed using Eq.~\ref{eq:Ci} and the following alternative formulation, which accounts only for depth error $Z$:
\begin{align}
\Tilde{C}^i = \left(1+\lVert \Delta \vZ^i \rVert_{\infty} \right)^{-1}.
\label{eq:Ci_Z}
\end{align}
Both confidence formulations produce a peak at the true depth, demonstrating their effectiveness. However, the confidence defined by Eq.~\ref{eq:Ci_Z} can also attain very high values at close candidate depths, sometimes even exceeding the confidence at the true depth and leading to incorrect predictions. In contrast, Eq.~\ref{eq:Ci} effectively suppresses the confidence at close candidate depths and prevents this failure mode.

\section{Hardware Assembly and CAD Models}
\input{sec/Supp_hardware}

\section{Calibration Procedure}

\input{sec/Supp_calibration}

\section{Additional Information on Data}

\subsection{Real-World Data}
\begin{figure}[h]
    \centering
    \includegraphics[width=1.0\linewidth]{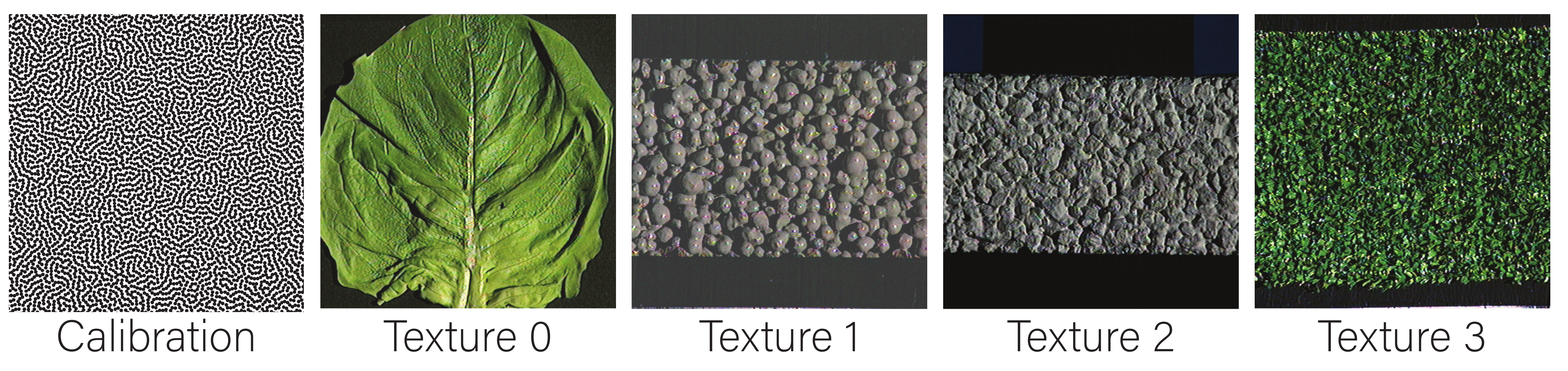}
    \caption{The planar textures utilized in our physical experiments. (Left) The high-contrast calibration pattern~\citeS{terao2020machine} used to align the dual-lens system and fit the continuous baseline parameter functions. (Right) The four distinct real-world patterns (Texture 0 to Texture 3) randomly sampled from the CUReT dataset~\cite{dana1999curet} to quantitatively evaluate the depth estimation accuracy of the prototype.}
    \label{fig:textures}
\end{figure}

To rigorously assess the performance of the proposed \dthreet~prototype, we utilize a set of distinct planar textures for system calibration and quantitative depth evaluation. As illustrated in Fig.~\ref{fig:textures}, a specific high-frequency calibration pattern from~\citeS{terao2020machine} is employed for all procedures detailed in the previous section, including vanishing point determination, stereo rectification, and baseline parameter tuning. 

For the experimental evaluation of our depth estimation accuracy, we capture front-parallel planes covered with four distinct patterns randomly selected from the CUReT dataset~\cite{dana1999curet}. See Fig.~\ref{fig:textures}. These patterns (denoted as Texture 0 through Texture 3) provide a diverse range of spatial frequencies and local image gradients. Testing across these varied surfaces ensures that the reported Mean Absolute Error (MAE) and working range metrics reflect robust, generalized performance across different visual target conditions rather than overfitting to a single ideal pattern.

To comprehensively evaluate the depth accuracy across a wide operating range, we collect two sequential datasets for each selected texture using our optical translation stages. The near-range dataset consists of 25 uniformly sampled depth planes ranging from 0.25\,m to 1.21\,m. The far-range dataset consists of an additional 25 uniformly sampled depth planes ranging from 1.01\,m to 1.97\,m. Together, these collections provide a dense, overlapping evaluation spectrum spanning from 0.25\,m to 1.97\,m, thoroughly validating the method's effectiveness across varied physical distances.

\subsection{Synthetic Data}

\begin{figure}[h]
    \centering
    \includegraphics[width=\linewidth]{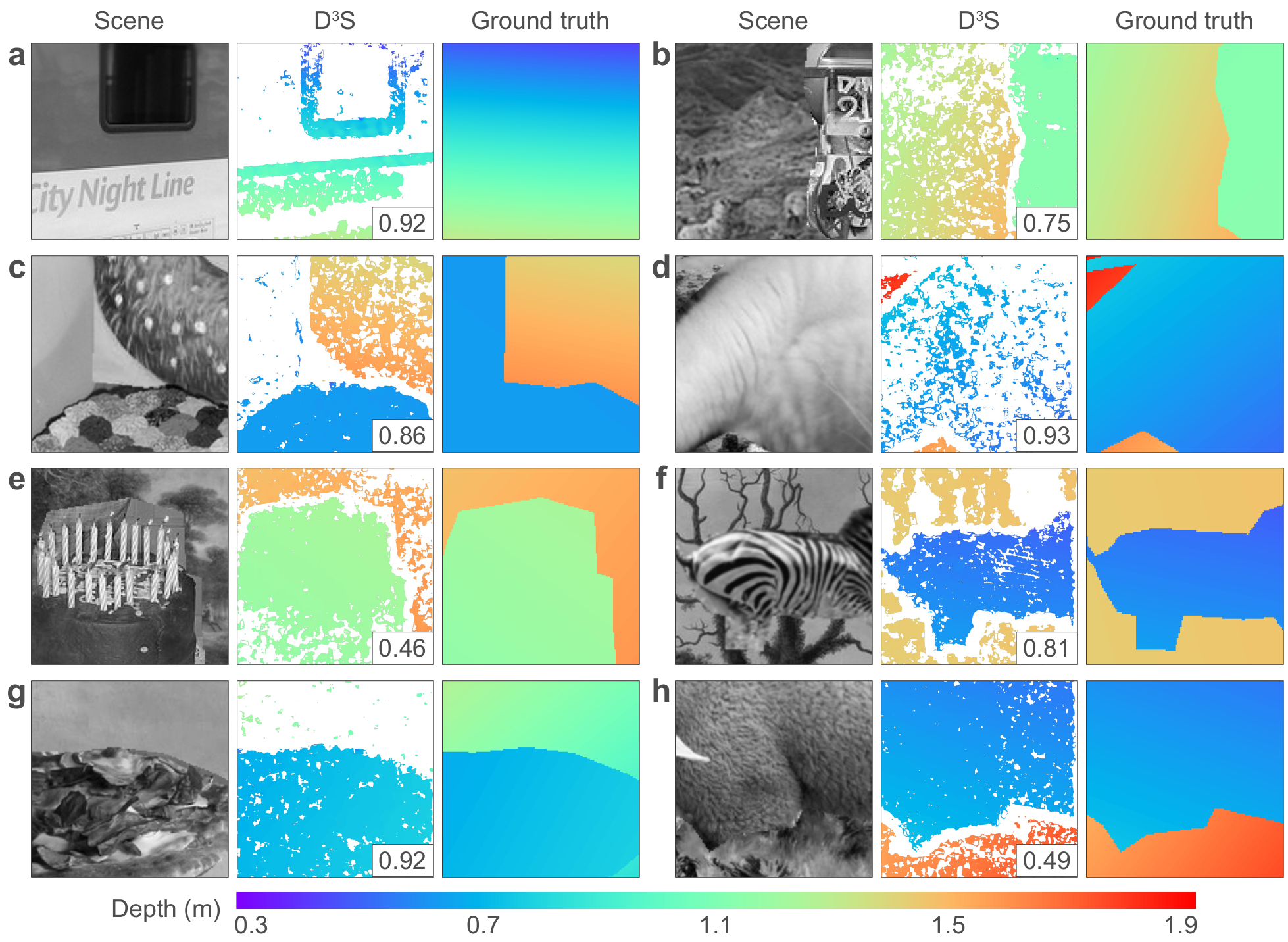}
    \caption{Sample synthetic scenes. These results are generated with a constant threshold $C_\text{thre} = 0.8$. Accurate depth estimation is maintained with depth boundaries and gradient-depth surfaces. The inserted numbers are MAE (cm) for depth estimation, calculated across valid pixels.}
    \label{fig:results_synthetic}
\end{figure}

We evaluate the simulated \dthreet~system on a synthetic dataset featuring realistic foreground and background textures, as well as complex depth discontinuities. To eliminate color bleeding artifacts at depth boundaries, we adopt a layer-wise rendering strategy for the foreground and background commonly used in previous works~\cite{guo2019compact,xu2025blurry}. The foregrounds and backgrounds are randomly sampled from the MS-COCO 2017 test set~\citeS{lin2014microsoft} and the Painting dataset~\citeS{crowley2014search}, respectively. Furthermore, a linear interpolation technique~\cite{guo2019compact} is applied to generate more natural tilted surfaces. Finally, we add Gaussian noise to the images. Each image in our synthetic dataset has a resolution of $1025 \times 1025$ pixels. Sample images and the corresponding depth estimation are shown in Fig.~\ref{fig:results_synthetic}.

\section{Specification of Comparison}

Table~\ref{tab:comparison_system} summarizes the complete specifications of the methods visualized in Fig.~\ref{fig:system_comparison}. Our system achieves a favorable combination of compactness and operating range, attaining a comparable working range while requiring a substantially smaller baseline and effective focal length (EFL). To further quantify this advantage, we estimate the end-point disparity error $\epsilon_d$ at the maximum depth $Z_\text{max}$ by solving Eq.~\ref{eq:epe} with $\epsilon_Z = 1$~cm and $Z = Z_\text{max}$. The resulting $\epsilon_d$ demonstrates a significant improvement over the other systems.

\input{sec/New_Table_system_comparison}


\section{Additional information of the \dthreet~Prototype}

\subsection{Point Spread Function}

\begin{figure}[h]
    \centering
    \includegraphics[width=1.0\linewidth]{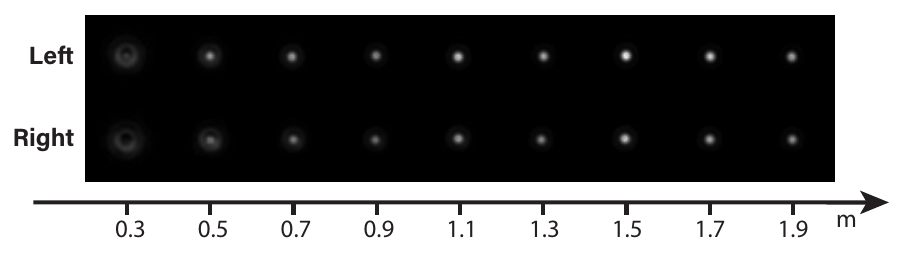}
    \caption{Visualization of the PSFs for the left and right sub-apertures of the \dthreet~prototype. The PSFs are captured at various depths from 0.3 to 1.9 m, illustrating the depth-dependent defocus blur utilized in our consensus formulation.}
    \label{fig:psf}
\end{figure}

To fully characterize the optical properties of the assembled \dthreet~prototype, we visualize the PSFs for both the left and right sub-apertures across the working range. Fig.~\ref{fig:psf} illustrates the PSFs captured from a point light source at discrete depth intervals ranging from 0.3 to 1.9 m. 


\subsection{Selection of Virtual Baselines}

In the $D^3S$ framework, the choice of the virtual baselines $\Delta \X$ dictates the magnitude of the sub-pixel shifts applied during the differential stereo matching process. To maximize depth accuracy, we treat the selection of these virtual baselines as a formal optimization problem over our calibration dataset $\mathcal{D}$. Specifically, we seek the set of $M$ virtual baselines $\{\Delta \X_m\}_{m=1}^M$ that minimizes the global Mean Absolute Error (MAE) between the predicted consensus depth $Z_{\text{pred}}$ and the ground truth depth $Z_{\text{gt}}$:

\begin{align}
\{\Delta \X_m^*\}_{m=1}^M = \arg\min_{\{\Delta \X_m\}} \frac{1}{|\mathcal{D}|} \sum_{i \in \mathcal{D}} \left| Z_{\text{pred}}^{(i)}(\{\Delta \X_m\}) - Z_{\text{gt}}^{(i)} \right|
\label{eq:virtual_baselines_argmin}
\end{align}

By evaluating this objective function across our captured near and far sequence depth planes, we identified $M=3$ as the optimal number of multi-baseline evaluations to balance computational efficiency with consensus robustness. For our specific hardware prototype and physical baseline ($\approx 3.84$\,mm), the optimal virtual baselines that minimize the calibration MAE were determined to be $0.45$\,mm, $0.50$\,mm, and $0.55$\,mm.



%% file: sec/Supp_full_experssion.tex
\subsection{Full Expression of the Depth Polynomial (Eq.~\ref{eq:general})}

By differentiating the image $I$ with respect to the simultaneous variation of optical parameters over time $t$, and solving for the depth $Z$, we obtain the cubic polynomial as listed in Eq.~\ref{eq:general}, $a_3 Z^3 + a_2 Z^2 + a_1 Z + a_0 = 0$. The exact expression of the coefficients $a_{0-3}$ are:

\begin{align}
a_0 &= |\Sigma|^2 s_0^2 A^2 \dot{Z} (\nabla^2_\Sigma I) \nonumber \\
a_1 &= |\Sigma|^2 s_0^2 (\nabla^2_\Sigma I) \left( -A^2 \rho \dot{Z} + \frac{A^2}{s}\dot{Z} - A \dot{A} \right) \nonumber \\
a_2 &= |\Sigma|^2 s_0^2 (\nabla^2_\Sigma I) \left( A^2 \dot{\rho} + \frac{A^2}{s^2}\dot{s} + 4A \dot{A} \rho - \frac{4A \dot{A}}{s} \right) - s_0 (I_{\vx}^\top \dot{\X}_0) - \dot{Z} (I_{\vx}^\top \vx) \nonumber \\
a_3 &= \dot{I} - |\Sigma|^2 s_0^2 (\nabla^2_\Sigma I) \left( A^2 \rho \dot{\rho} + \frac{A^2 \rho}{s^2}\dot{s} - \frac{A^2}{s}\dot{\rho} - \frac{A^2}{s^3}\dot{s} + A \dot{A} \rho^2 - \frac{2A \dot{A}}{s}\rho + \frac{A \dot{A}}{s^2} \right)
\label{eq:coeffs}
\end{align}
To solve for $Z$ directly, we normalize the polynomial by dividing by the leading coefficient $a_3$, yielding $Z^3 + b Z^2 + c Z + d = 0$, where $b = \frac{a_2}{a_3}$, $c = \frac{a_1}{a_3}$, and $d = \frac{a_0}{a_3}$. While a cubic polynomial generally yields three roots, the physical depth $Z$ must be a real, positive scalar. By omitting the complex cube roots of unity ($\omega$ and $\omega^2$) required for the secondary and tertiary roots, Cardano's formula allows us to isolate the primary real root directly, defining the intermediate variables in Eq.~\ref{eq:cardano_vars} to yield the final depth in Eq.~\ref{eq:cardano_z}:

\begin{align}
p &= c - \frac{b^2}{3}, \quad q = \frac{2b^3}{27} - \frac{bc}{3} + d, \quad \Delta = \left(\frac{q}{2}\right)^2 + \left(\frac{p}{3}\right)^3 \label{eq:cardano_vars} \\
Z &= -\frac{b}{3} + \sqrt[3]{-\frac{q}{2} + \sqrt{\Delta}} + \sqrt[3]{-\frac{q}{2} - \sqrt{\Delta}} \label{eq:cardano_z}
\end{align}

To understand the nature of the omitted roots, let $u$ and $v$ represent the two primary real cube roots from Eq.~\ref{eq:cardano_z}, such that the isolated real root is $-\frac{b}{3} + u + v$. The remaining two roots evaluate to $-\frac{b}{3} - \frac{1}{2}(u+v) \pm i\frac{\sqrt{3}}{2}(u-v)$. For realistic physical camera configurations, the discriminant is typically positive ($\Delta > 0$). This guarantees that $\sqrt{\Delta}$ is real and $u \neq v$, meaning the imaginary component $\pm i\frac{\sqrt{3}}{2}(u-v)$ is strictly non-zero. Consequently, the two omitted roots are complex conjugates and mathematically invalid for physical depth. In rarer cases where multiple real roots exist ($\Delta \leq 0$), the principal root $Z$ extracted by Eq.~\ref{eq:cardano_z} naturally corresponds to the physically valid depth within the camera's operable working range.

\subsection{Special Cases of the Depth Equation}

When specific optical parameters are isolated (or varied linearly without axial camera movement, i.e., $\dot{Z} = 0$), the complex general relationship simplifies significantly. The exact closed-form solutions for these special cases are summarized in Table~\ref{tab:special_cases}.

Our solutions complements ones in Alexander~\cite{alexander2019theory}. The fundamental difference between the two lies in the handling of spatial magnification. Our $D^3$ approach scales the images into uniform magnification through a scaling factor ($s_0$) prior to differentiation, avoiding an additional differential term $I_{\vx}^\top \vx$. In contrast, Alexander's framework differentiates the raw image directly. When the varying parameter is $\X_0$, $\rho$, or $A$, the terms $s$ and $s_0$ are constant. That is, the $D^3$ approach is mathematically identical to Alexander's when $s=s_0$. 

However, when evaluating the combined parameter variation ($\dot{\X}_0, \dot{\rho}, \dot{s} \neq 0$), Alexander's Partial Differential Equation natively couples $\dot{s}$ with spatial shifts, becoming quadratic with respect to inverse depth: $\dot{I} = |\Sigma|^2 A^2 (\nabla^2_\Sigma I) \big[ s\dot{s}(\frac{1}{Z} - \rho)^2 + (\dot{s} - s^2\dot{\rho})(\frac{1}{Z} - \rho) - s\dot{\rho} \big] + \frac{s}{Z} (I_{\vx}^\top \dot{\X}_0) - \frac{\dot{s}}{s} (I_{\vx}^\top \vx)$. Consequently, a single linear solution does not exist in their framework for this case on a per-pixel basis, yielding multiple possible solutions for Alexander depending on the roots of the quadratic polynomial. While Alexander requires solving the constraint across a patch of pixels to resolve this mathematical ambiguity, our magnification correction successfully decouples the spatial shifts, enabling a direct, closed-form per-pixel computation.

\begin{table}[h]
\centering
\caption{Special cases of the depth equation. Comparison between our magnification-corrected formulation ($D^3$) and the raw-image formulation by Alexander~\cite{alexander2019theory}. For the combined variation case, Alexander's framework yields a quadratic polynomial requiring spatial patches to resolve the mathematical ambiguity, whereas our $D^3$ correction achieves a linear, per-pixel closed-form solution.}
\label{tab:special_cases}
\resizebox{\textwidth}{!}{
\begin{tabular}{@{}lcc@{}}
\toprule
Varying Parameter & Our Method ($D^3$) & Alexander~\cite{alexander2019theory} \\
\midrule
Lateral Translation ($\X_0$) & \multicolumn{2}{c}{$Z = \frac{s_0 (I_{\vx}^\top \dot{\X}_0)}{\dot{I}}$} \\
Optical Power ($\rho$) & \multicolumn{2}{c}{$Z = \left[ \rho - \frac{1}{s} - \frac{\dot{I}}{|\Sigma|^2 s_0^2 A^2 \dot{\rho} \nabla^2_\Sigma I} \right]^{-1}$} \\
Aperture Radius ($A$) & \multicolumn{2}{c}{$Z = \left[ \rho - \frac{1}{s} \pm \frac{1}{s} \sqrt{\frac{\dot{I} s^2}{|\Sigma|^2 s_0^2 A \dot{A} \nabla^2_\Sigma I}} \right]^{-1}$} \\
Sensor Distance ($s$) & $Z = \left[ \rho - \frac{1}{s} - \frac{s^2 \dot{I}}{|\Sigma|^2 s_0^2 A^2 \dot{s} \nabla^2_\Sigma I} \right]^{-1}$ & $Z = \left[ \rho - \frac{1}{s} - \frac{s \left( \dot{I} - \frac{\dot{s}}{s}(I_{\vx}^\top \vx) \right)}{|\Sigma|^2 A^2 \dot{s} \nabla^2_\Sigma I} \right]^{-1}$ \\
Combined ($\X_0, \rho, s$) & $Z_{D^3} = \frac{s_0 (I_{\vx}^\top \dot{\X}_0) - |\Sigma|^2 s_0^2 A^2 \left(\dot{\rho} + \frac{\dot{s}}{s^2}\right) (\nabla^2_\Sigma I)}{\dot{I} - |\Sigma|^2 s_0^2 A^2 \left(\dot{\rho} + \frac{\dot{s}}{s^2}\right) (\nabla^2_\Sigma I) \left(\rho - \frac{1}{s}\right)}$ & $Z = \left[ \frac{-c_1 \pm \sqrt{c_1^2 - 4c_2c_0}}{2c_2} \right]^{-1}$ \\
\midrule
\multicolumn{3}{l}{
\begin{tabular}{@{}l@{}}
\small \textit{Remark:} The explicit coefficients for Alexander's \textbf{Combined} quadratic polynomial ($c_2 Z^{-2} + c_1 Z^{-1} + c_0 = 0$) evaluate to: \\
$\begin{aligned}
c_2 &= |\Sigma|^2 A^2 s \dot{s} (\nabla^2_\Sigma I) \\
c_1 &= |\Sigma|^2 A^2 (\nabla^2_\Sigma I) \left( \dot{s} - 2s\dot{s}\rho - s^2\dot{\rho} \right) + s(I_{\vx}^\top \dot{\X}_0) \\
c_0 &= |\Sigma|^2 A^2 (\nabla^2_\Sigma I) \left( s\dot{s}\rho^2 - \dot{s}\rho + s^2\rho\dot{\rho} - s\dot{\rho} \right) - \frac{\dot{s}}{s}(I_{\vx}^\top \vx) - \dot{I}
\end{aligned}$ \\
\end{tabular}
} \\
\bottomrule
\end{tabular}
}
\end{table}

\paragraph{Axial Translation ($\dot{Z} \neq 0$):}
Unlike the parameters above, pure axial translation does not yield a simple closed-form solution. Let $\delta = \frac{1}{Z} - \rho$. The relationship can be expressed in its factored form, as defined in Eq.~\ref{eq:axial_translation}:

\begin{align}
\dot{I} = \dot{Z}\frac{1}{Z}(I_{\vx}^\top \vx) - |\Sigma|^2 (\nabla^2_\Sigma I) \frac{s_0^2 A^2}{s} \dot{Z} (1+s\delta)(\delta+\rho)^2
\label{eq:axial_translation}
\end{align}

%% file: sec/Supp_hardware.tex


\begin{figure}[t]
    \centering
    \includegraphics[width=0.7\linewidth]{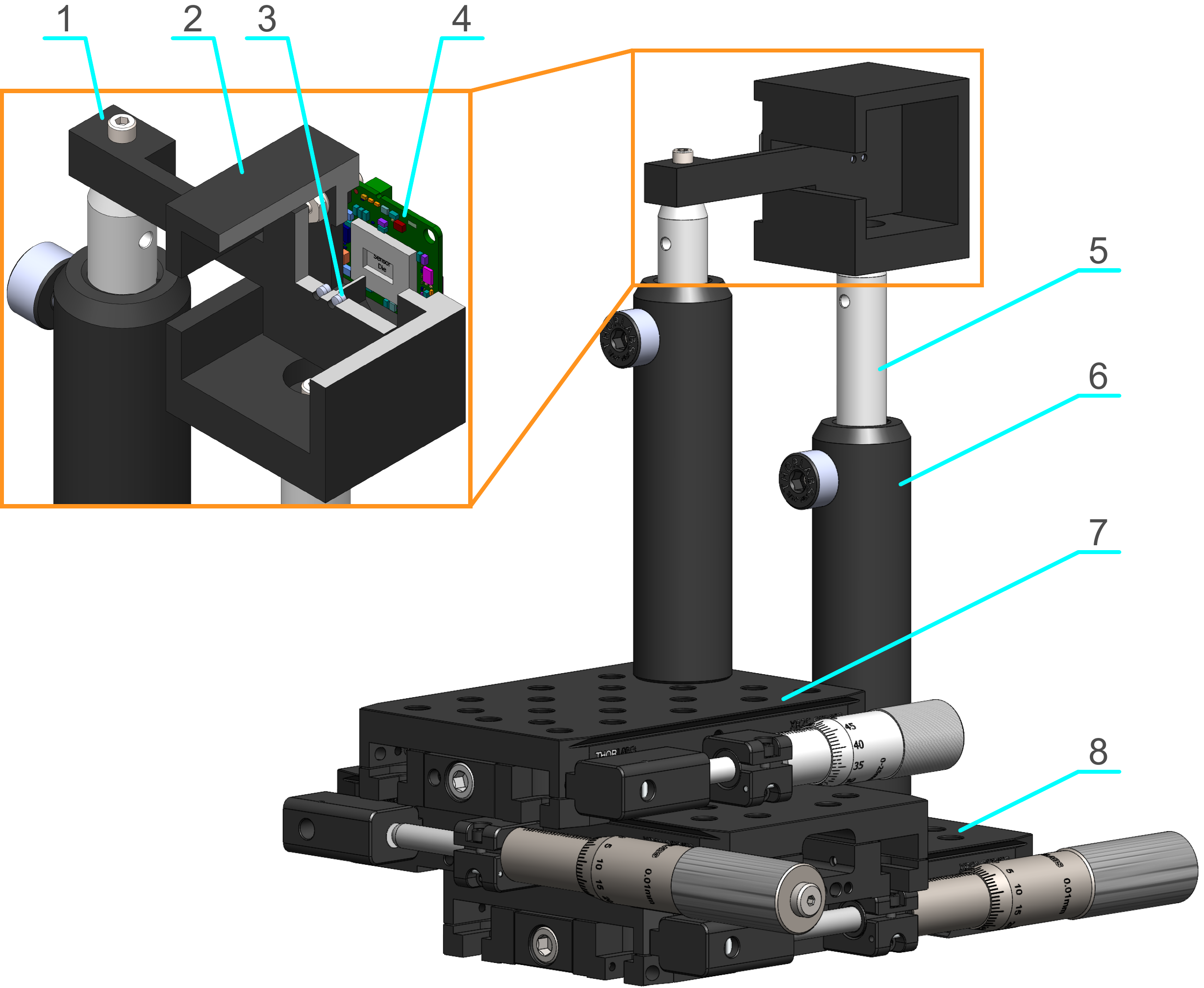}
    \begin{tabular}{@{}lccc@{}}
\toprule
No. & Component & Type/Description & Quantity \\
\midrule
1 & Lens Mount & Custom 3D-Printed Dual-Lens Mount & 1 \\
2 & Housing & Custom 3D-Printed Camera Housing & 1 \\
3 & Lens & \makecell[c]{2\,mm Dia. $\times$ 12\,mm FL, VIS 0$^\circ$ Coated,\\Achromatic Lens} & 2 \\
4 & Camera & \makecell[c]{Basler dart daA3840-45um\\(Bare-board/No-Mount)} & 1 \\
5 & Optical Post & $\varnothing$1/2" Optical Post, SS, L = 3" & 2 \\
6 & Post Holder & $\varnothing$1/2" Post Holder, Spring-Loaded, L = 3" & 2 \\
7 & Short Linear Stage & 25\,mm Travel Linear Translation Stage & 1 \\
8 & Long Linear Stage & 50\,mm Travel Linear Translation Stage & 2 \\
\bottomrule
\end{tabular}
    \caption{CAD rendering of the \dthreet{} Consensus prototype (Fig.~\ref{fig:prototype}b). The labeled components correspond to the parts listed in the accompanying table.}
    \label{fig:CAD}
\end{figure}


Figure~\ref{fig:CAD} lists the details of all hardware components used to build the \dthreet~prototype. We want to emphasize that the physical prototype we presented represents just one possible hardware implementation of the \dthreet~Consensus mechanism. People can readily reproduce our approach using different hardware solutions. 


The physical prototype is assembled on the optical stages following these steps:


\begin{enumerate}
    \renewcommand{\labelenumi}{\Roman{enumi}.}
    \item \textbf{Lens Preparation:} Insert the two achromatic lenses (\textbf{3}) into the custom 3D-printed dual-lens mount (\textbf{1}) to form the dual-lens optical plate.
    \item \textbf{Sensor Mounting:} Secure the bare-board sensor (\textbf{4}) inside the custom 3D-printed camera housing (\textbf{2}) using four mounting screws. 
    \item \textbf{Housing Positioning:} Attach the camera housing (\textbf{2}) to a 3-inch optical post (\textbf{5}). Insert this post into a post holder (\textbf{6}) and mount it onto the far-left hole of the primary 50\,mm (long) linear translation stage (\textbf{8}).
    \item \textbf{Stage Stacking:} Stack the 25\,mm (short) linear translation stage (\textbf{7}) on top of the second 50\,mm (long) linear translation stage (\textbf{8}) to form an X-Y positioning mechanism. Mount this assembled stack on the right side of the primary long linear stage (\textbf{8}).
    \item \textbf{Lens Mount Attachment:} Attach the assembled lens mount (\textbf{1}) to the second optical post (\textbf{5}). Secure this post inside the second post holder (\textbf{6}) and mount it on top of the stacked X-Y linear stages (\textbf{7}, \textbf{8}). 
    \item \textbf{Axial Alignment (Focus):} Use the primary long linear stage (\textbf{8}) to carefully position the lens mount (\textbf{1}) inside the camera housing (\textbf{2}), ensuring that the partition on the lens mount faces the sensor (\textbf{4}). Set the initial sensor-to-lens distance to approximately 13\,mm. Place a calibration target at a distance of 1\,m, and translate the lens mount axially until the image appears sharp.
    \item \textbf{Lateral Alignment (Separation):} Finally, fine-tune the stacked X-Y linear stages (\textbf{7}, \textbf{8}) to align the optical partition such that the separation line is perfectly centered between the dual images captured by the sensor (\textbf{4}).
\end{enumerate}


%% file: sec/Supp_calibration.tex




\subsection{Anchor Points Determination} 
\label{secsec:anchor}
First, we determine the pixel coordinates $(\x_0^\infty, \x_1^\infty)$ in the two images $I_0$ and $I_1$ that correspond to a point located at infinite depth. These coordinates serve as anchors for cropping the raw measurements to produce the aligned images $I_0$ and $I_1$. With this alignment, an infinitely distant point yields zero disparity between the two cropped images.

To estimate these anchor points, we exploit the fact that the projection of an infinitely distant point remains stationary when a planar target is translated axially. We therefore capture a sequence of $N$ images while moving a textured calibration target to different distances $\{Z^j\}_{j=1}^N$. The calibration texture used in this procedure is shown in Fig.~\ref{fig:textures}.

For a candidate window centered at pixel coordinate $\vx$, we measure the local translational shift between consecutive frames using SIFT keypoint matching followed by RANSAC-based homography estimation. The true vanishing point $\vx_i^\infty$ is then identified as the coordinate that minimizes the total squared shift across the entire image sequence:
\begin{align}
\vx_{i}^\infty =
\arg\min_{\vx}
\sum_{j=1}^{N}
\mathcal{F}_{\text{shift}}\left(\vx; I_i(Z^j), I_i(Z^{j+1})\right), i \in \{0,1\}
\label{eq:vp_argmin}
\end{align}
where $\mathcal{F}_{\text{shift}}$ denotes the mean squared translation of SIFT keypoints within the window centered at $\vx$ between images $I(Z^j)$ and $I(Z^{j+1})$. This optimization is performed independently to locate the vanishing points for $I_0$ and $I_1$.

\subsection{Rectification}

After cropping the $I_0$ and $I_1$ using the anchor points, we further rectify $I_0$ and $I_1$ to ensure that corresponding features only require a horizontal translation to be matched. For this procedure, we utilize the exact same axial-motion calibration dataset captured in Sec.~\ref{secsec:anchor} and perform the following steps:

\begin{enumerate}
    \renewcommand{\labelenumi}{\Roman{enumi}.}
    \item \textbf{Feature Matching:} We extract Oriented FAST and Rotated BRIEF (ORB) features from both sub-images across our calibration sequence and establish robust correspondences using a $K$-nearest neighbors ratio test.
    \item \textbf{Transformation Estimation:} Using these matched keypoints, we estimate a 2D similarity transformation (incorporating uniform scale, rotation, and 2D translation) for each individual frame via RANSAC.
    \item \textbf{Global Consensus:} To ensure a stable and globally consistent alignment across the entire working range, we decompose the transformations from all calibration frames and compute the median scale ($\chi$), median rotation angle ($\theta$), and median vertical translation ($\Delta y$).
    \item \textbf{Image Warping:} We construct a constant geometric transformation matrix $H_{\text{rect}}$ using these median values. Crucially, we intentionally enforce zero horizontal translation ($\Delta x = 0$) in this matrix to perfectly preserve the physical horizontal disparity caused by the dual-lens baseline:
    
    \begin{align}
    H_{\text{rect}} = \begin{bmatrix} \chi \cos \theta & -\chi \sin \theta & 0 \\ \chi \sin \theta & \chi \cos \theta & \Delta y \\ 0 & 0 & 1 \end{bmatrix}
    \label{eq:rectification_matrix}
    \end{align}
    
    Warping the right sub-image with $H_{\text{rect}}$ effectively compensates for any physical roll, pitch, or scale discrepancies introduced by manufacturing tolerances in the 3D-printed mount, yielding a perfectly rectified stereo pair where epipolar lines are strictly horizontal. 
\end{enumerate}

\subsection{Parameter tuning.} 
With the images rectified, the final step is to calibrate the baseline-dependent scale multiplier $a(\Delta \X)$ and depth offset $b(\Delta \X)$ in Eq.~\ref{eq:d3_actual} so that the \dthree{} model accurately reflects the optical characteristics of our physical sensor and lenses. We use the same calibration dataset described in Sec.~\ref{secsec:anchor} and perform the following procedure.

\begin{enumerate}
    \item \textbf{Parameter Optimization:}
    For a set of sampled virtual baselines $\{\Delta \X^k\}$, we determine the optimal parameters $a(\Delta \X^k)$ and $b(\Delta \X^k)$ by minimizing the Mean Absolute Error (MAE) between the predicted depth and the ground truth over the calibration dataset:
    \begin{align}
        a(\Delta \X^k), b(\Delta \X^k)
        = \arg\min_{a,b}
        \sum_{j=1}^{N}
        \left\Vert
        Z_{\text{\dthree}}\!\left(I_0(Z^j), I_1(Z^j); a,b,\Delta\X^k \right)
        - Z^j
        \right\Vert.
    \end{align}
    Solving this objective yields the optimal scale multiplier and depth offset for each sampled baseline $\Delta \X^k$.

    \item \textbf{Polynomial Regression:}
    The discrete parameter values obtained above are then used to estimate continuous parameter functions through third-order polynomial regression. We select the third-order polynomial empirically after viewing the trend of the sampled points. Let $\Delta\X$ denote the virtual baseline. The calibrated functions take the form
    \begin{align}
    a(\Delta\X) &= c_{a,3}\lVert\Delta\X\rVert^3 + c_{a,2}\lVert\Delta\X\rVert^2 + c_{a,1}\lVert\Delta\X\rVert + c_{a,0}, \label{eq:param_a} \\
    b(\Delta\X) &= c_{b,3}\lVert\Delta\X\rVert^3 + c_{b,2}\lVert\Delta\X\rVert^2 + c_{b,1}\lVert\Delta\X\rVert + c_{b,0}, \label{eq:param_b}
    \end{align}
    where the coefficients $c_{a,i}$ and $c_{b,i}$ are obtained from regression over the calibration dataset. For an arbitrary virtual baseline $\Delta \X$, the parameters $a(\Delta\X)$ and $b(\Delta\X)$ are computed by evaluating these polynomials.
\end{enumerate}

\begin{figure}[h]
    \centering
    \includegraphics[width=0.5\linewidth]{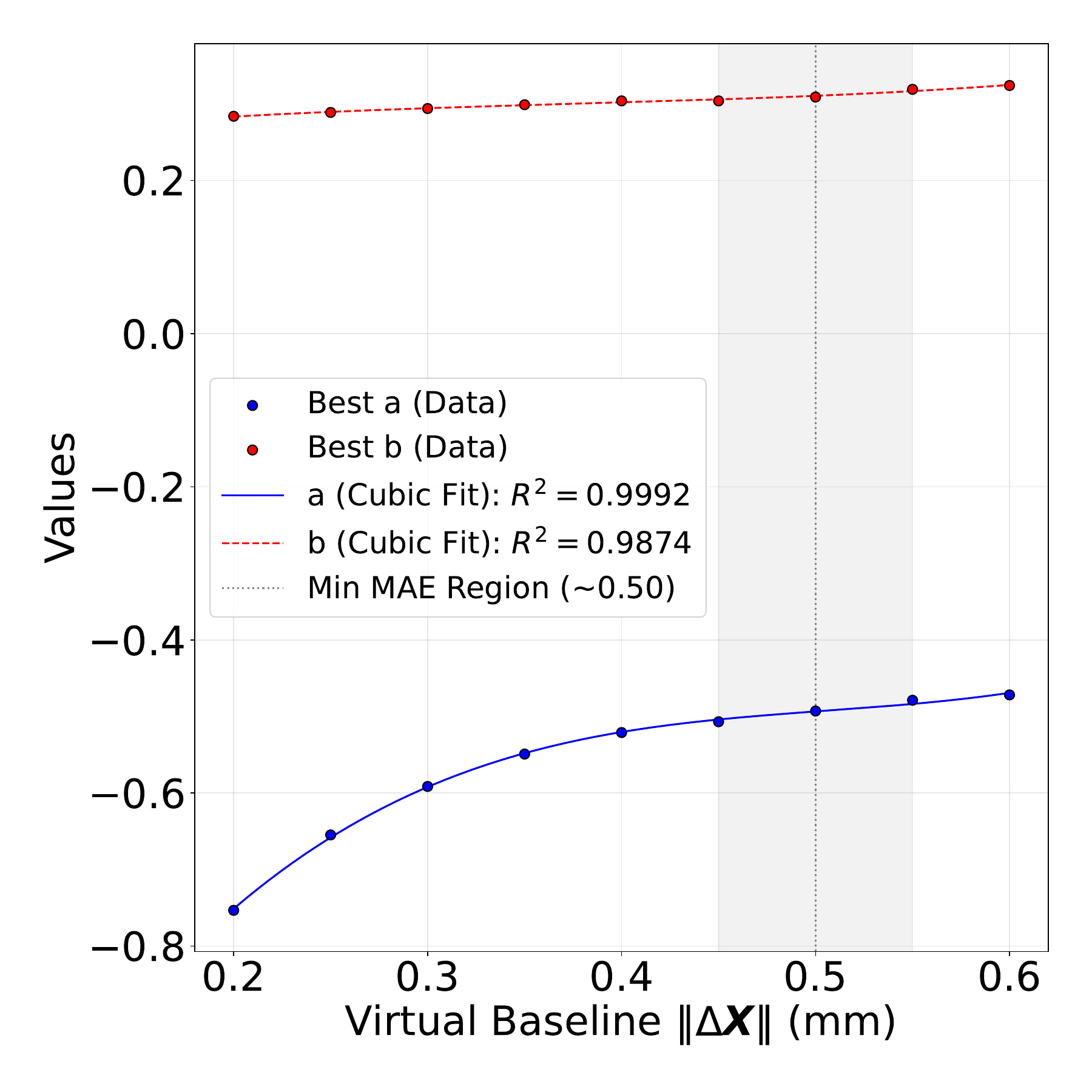}
    \caption{Calibration of the baseline-dependent parameters $a(\Delta \X)$ and $b(\Delta \X)$. The discrete points indicate the optimal scale multiplier $a(\Delta\X)$ (blue) and depth offset $b(\Delta\X)$ (red) determined via optimization for each virtual baseline $\Delta\X$. The continuous curves illustrate the robust third-order polynomial regression (cubic fit) applied to these discrete values, both exhibiting high coefficients of determination ($R^2 > 0.98$). The shaded gray area denotes the minimum MAE region from which our final virtual baselines ($0.45$\,mm, $0.50$\,mm, and $0.55$\,mm) were selected for depth estimation. }
    \label{fig:ab_fit}
\end{figure}

%% file: sec/New_Table_system_comparison.tex
\begin{table}[t!]
  \caption{Simulation comparison with prior depth-from-defocus and dual-defocus stereo systems. The notations $\lVert \Delta \X\rVert$, $s_0$, $p$, $\epsilon_d$ indicate the baseline, EFL, pixel pitch, and end-point-error in disparity. The max depth $Z_\text{max}$ denotes the upper bound of working range with MAE < 1 cm.  Despite using a significantly shorter baseline and EFL, our system achieves comparable working range to most others. This indicates that the proposed system can achieve a significantly higher accuracy and longer working range if under uniform form factors. }
  \label{tab:comparison_system}
  \centering
  \setlength{\tabcolsep}{1pt}
  \begin{tabular}{@{}lcccccc@{}}
\toprule
\makecell[l]{Method} &
\makecell[c]{Venue} &
\makecell[c]{$\lVert\Delta\X\rVert$\\(mm) $\downarrow$} &
\makecell[c]{$s_0$\\(mm) $\downarrow$} &
\makecell[c]{$p$\\(\textmu m)} &
\makecell[c]{$Z_\text{max}$ (m)\\<1 cm MAE $\uparrow$} &
\makecell[c]{$\epsilon_{d}$ @ $Z_\text{max}$\\(px) $\downarrow$} \\
\midrule
Takeda \etal~\cite{takeda2012coded}\textsuperscript{\dag\ddag} & VISAPP \textquotesingle12 & 14.0 & 30.0 & 6.11 & 0.37 & 5.08 \\
Takeda \etal~\cite{takeda2013fusing}\textsuperscript{\dag\ddag} & CVPR \textquotesingle13 & 14.0 & 51.3 & 6.11 & 1.47 & 0.54 \\
Tang \etal~\cite{tang2017depth}\textsuperscript{\dag$\|$} & CVPR \textquotesingle17 & 1.72 & 4.12 & 1.55 & 0.36 & 0.36 \\
Tan \etal~\cite{tan2021codedstereo}\textsuperscript{\dag\S} & CVPR \textquotesingle21 & 22.0 & 52.6 & 4.80 & No satisfied range & -- \\
Lopez \etal~\cite{lopez2024low}\textsuperscript{\dag\S} & ECCVW \textquotesingle24 & 120 & 100 & 3.72 & 5.83 & 0.95 \\
Liu \etal~\cite{liu2025learned}\textsuperscript{\dag\S} & CVPR \textquotesingle25 & 75.0 & 35.0 & 5.86 & 1.92 & 1.21 \\
Ou \etal~\cite{ou2025learning}\textsuperscript{\dag\P} & Vs. Cp. \textquotesingle25 & 70.0 & 35.9 & 5.86 & 1.78 & 1.35 \\
\textbf{Ours} & -- & 3.84 & 12.1 & 2.00 & 1.56 & \textbf{0.11} \\
\bottomrule
\end{tabular}\\
\scriptsize \dag Converted from reported EPE from the original paper. 
\ddag Three simulated step scenes. \S Scene Flow dataset~\citeS{mayer2016large}. \P InStereo2k dataset~\citeS{bao2020instereo2k}. $\|$Middlebury Stereo 2006 dataset~\citeS{hirschmuller2007evaluation}. \cite{crowley2014search}
\end{table}